\documentclass[10pt]{article}
\usepackage[margin=0.9in]{geometry}
\usepackage{epsfig,amssymb,soul}
\usepackage[normalem]{ulem}

\usepackage[breaklinks=true,colorlinks=true,hypertexnames=false,linktoc=section,pdfencoding=auto,psdextra=true,linkcolor=black,anchorcolor=black,citecolor=black,filecolor=black,menucolor=black,runcolor=black,urlcolor=black]{hyperref}
\usepackage{graphicx}
\usepackage{epstopdf}
\usepackage{amsfonts,amsmath}
\usepackage{bm}
\usepackage{setspace}
\usepackage{comment}
\usepackage{siunitx}
\usepackage{mathrsfs}
\usepackage{multirow}
\usepackage{subcaption}
\usepackage[hang,flushmargin,bottom]{footmisc}
\usepackage{mathtools}
\usepackage{lipsum}
\usepackage{array}
\usepackage{enumerate}
\usepackage{ifthen}
\usepackage[affil-it]{authblk}
\usepackage{cite}
\usepackage{leftidx}
\usepackage{relsize}
\usepackage{braket}
\usepackage{float}
\usepackage[left]{lineno}

\usepackage[dvipsnames]{xcolor}

\newcommand{\etal}{\textit{et al}.~}
\providecommand{\keywords}[1]{\textbf{\textit{Keywords-}} #1}

\title{Fractional-Order Shell Theory: Formulation and Application to the Analysis of Nonlocal Cylindrical Panels}
\author[1]{Sai Sidhardh}
\author[1]{Sansit Patnaik}
\author[1]{Fabio Semperlotti \thanks{All correspondence should be addressed to \textit{fsemperl@purdue.edu}
}}
\affil[1]{Ray W. Herrick Laboratories, School of Mechanical Engineering, Purdue University, West Lafayette, Indiana, USA - 47907}
\date{}

\begin{document}
\maketitle
\vspace*{-1cm}
\begin{abstract}
We present a theoretical and computational framework based on fractional calculus for the analysis of the nonlocal static response of cylindrical shell panels. The differ-integral nature of fractional derivatives allows an efficient and accurate methodology to account for the effect of long-range (nonlocal) interactions in curved structures. More specifically, the use of frame-invariant fractional-order kinematic relations enables a physically, mathematically, and thermodynamically consistent formulation to model the nonlocal elastic interactions. In order to evaluate the response of these nonlocal shells under practical scenarios involving generalized loads and boundary conditions, the fractional-Finite Element Method (f-FEM) is extended to incorporate shell elements based on the first-order shear-deformable displacement theory. 
Finally, numerical studies are performed exploring both the linear and the geometrically nonlinear static response of nonlocal cylindrical shell panels. This study is intended to provide a general foundation to investigate the nonlocal behavior of curved structures by means of fractional order models. 
\end{abstract}

\noindent\keywords{Fractional calculus, Nonlocal shells, Cylindrical panels, Geometric nonlinearity}
\section{Introduction}
Shell panels and, more in general, curved structures are ubiquitous in the design of lightweight structural applications. Several example can be found throughout the aerospace, naval, and civil engineering sectors. Their prevalence in several technical areas has long fueled research on design and modeling aspects of shell structures\cite{librescu1975elastostatics,amabili2003review,carrera2003historical,alijani2014non}. 
Recent theoretical and experimental investigations have highlighted the existence of size-effects or, equivalently, nonlocal effects in several classes of structures. For example, size-effects within low-dimensional (micro- and nano-scale) structures, such as carbon nanotubes, emerge due to the long-range forces that become more prevalent at these scales \cite{arash2012review,behera2017recent}. Several studies have also demonstrated the presence of nonlocal interactions at the macro-scale in complex structures like porous solids \cite{patnaik2021porous,bulle2021human,fellah2004ultrasonic}, periodic structures \cite{russillo2022wave,nair2021nonlocal}, intentionally engineered nonlocal structures \cite{zhu2020nonlocal}, and sandwiched structures \cite{romanoff2020design}. The inability of the classical (local) continuum theory to capture the nonlocal effects has been one of the main drivers fostering the development of nonlocal continuum theories.

To-date, several classes of constitutive models have been developed. Broadly speaking, the classical nonlocal models developed so far can be classified in strain-driven \cite{eringen1972linear} and stress-driven \cite{romano2017stress} approaches. Seminal studies, undertaken by Kr\"oner\cite{kroner1967elasticity} and Eringen \cite{eringen1972linear}, proposed integral models where the nonlocal behavior was captured within stress-strain constitutive relations. The integral approach proposed in this studies can be classified as a strain-driven approach, wherein the stress at a point is expressed via a convolution of the strain field scaled by an attenuation kernel \cite{polizzotto2001nonlocal}. This convolution integral extends over the domain of nonlocal influence. This specific approach gained rapidly traction within the research community owing to a clear and intuitive representation of the nonlocal interactions using convolution integrals. The numerical complexity involved in the practical simulation of the strain-driven approach led to the development of the differential model of nonlocal elasticity, which was derived from the integral theory by the use of exponential kernels and was more amenable to numerical simulations \cite{eringen1983differential}. Several numerical studies, mostly focused on beam and plate elements \cite{arefi2020electro,aminipour2020analysis,babaei2021nonlinear}, followed the introduction of this approach. The studies dedicated to the modeling of nonlocal effects in shell structures focused, for the most part, on carbon nanotubes \cite{lu2007application,reddy2008nonlocal,arash2012review}. While a common finding in these studies was the the softening effect due to the presence of nonlocal interactions, the differential theories also highlighted the occurrence of an apparently paradoxical physical behavior. In other terms, it was observed that, for selected choices of the loading and boundary conditions\cite{challamel2014nonconservativeness}, nonlocal effects vanished. This inconsistent behavior was attributed to the ill-posed nature of the governing equations\cite{romano2017stress} that did not guarantee a positive-definite formulation for the deformation energy density. In order to address the inconsistencies mentioned above, a stress-driven approach was developed by Romano \etal \cite{romano2017stress} which derived additional constraints, in the form of nonlocal constitutive boundary conditions, to achieve a well-posed formulation. In this approach, the strain at a given point in the material is expressed via the convolution of the stress field, scaled by an attenuation kernel, at all points within the nonlocal horizon of influence. The stress-driven approach to nonlocal elasticity has also been utilized in studies on beams\cite{romano2017stress}, curved beams\cite{zhang2020exact}, and tubular structures\cite{malikan2020free}. However, the analytical complexities involved in the implementation of the stress-driven approach have, so far, prevented its application to higher-dimensional structures (for example, plates and shells) subjected to generalized external loads. The above summary highlights that, while nonlocal models of elasticity have seen a rapid expansion in recent years, there are still important obstacles on the way to achieve consistent theories under generalized loading and boundary conditions \cite{batra159misuse}; this consideration is even more true if one restricts the assessment to available theories to model the nonlocal response of curved panels. 

In recent years, fractional order models of elasticity have emerged as a powerful alternative to efficiently and accurately simulate nonlocal response of materials and structures \cite{shitikova2021fractional,lazopoulos2006non,atanackovic2009generalized,di2013mechanically,sumelka2014thermoelasticity,sumelka2014fractional,alotta2017finite}. In particular, Patnaik \etal\cite{patnaik2019generalized,patnaik2019FEM} proposed a fractional-order constitutive theory for nonlocal solids with fractional-order strain-displacement relations. The differ-integral nature of the fractional derivatives makes them a suitable alternative to the integral type relations commonly encountered in the studies on nonlocal elasticity. This class of constitutive relations falls under the general category of a displacement-driven approach to nonlocal elasticity with nonlocal kinematic relations\cite{patnaik2022displacement}. It has been established that this displacement-driven approach to modeling nonlocal interactions provides well-posed governing equations with self-adjoint linear operators following from a physically consistent positive-definite definition for deformation energy density\cite{patnaik2019FEM}. More importantly, unlike other existing nonlocal theories described above, the proposed fractional-order constitutive model satisfies the thermodynamic balance laws in a rigorous manner\cite{sidhardh2021thermodynamics}. Further, by means of variational methodologies available for the fractional-order models, thanks to the energy framework, finite element models may be developed for a numerical solution of fractional-order boundary value problems\cite{patnaik2019FEM}. Employing this framework, a consistent softening effect of long-range interactions has been documented over the elastic response of nonlocal beams and plates\cite{patnaik2019FEM,sidhardh2020geometrically,patnaik2020geometrically,sidhardh2021fractional}. The coherence in observations of softening caused by long-range interactions across a wide range of studies attests to the suitability of fractional-order models for nonlocal elasticity. However, all of the above studies over fractional-order theories for nonlocal elasticity are restricted to beams and plates and, more in general, structures without curvature. To the best of the author's knowledge, there is currently no available fractional-order theory to model the nonlocal response of shell structures. 

In concluding this introductory section, we would like to mention an important aspect the affects the development of the fractional order shell theory.
Previous fractional order models (e.g. for beams and plates) were developed by making use of orthonormal Cartesian coordinate axes. Unlike these cases, employing Cartesian coordinates for curved structures would require a complete 3D theoretical and numerical analysis. While this can be undertaken for simple cases, a generalized shell theory would be required for the development of a reduced-order model for curved structures. Therefore, it may be concluded that the modeling of curved structures requires curvilinear coordinate axes. Sufficient literature is available for a study on the local elastic response of curved structures using integer-order models for shell theories\cite{reddy2006theory}. However, the gap in a similar literature for nonlocal elastic structures is clear. This may be attributed to a lack of the mathematical literature and associated methodologies or the use of fractional calculus in curvilinear coordinate systems, which appear not to be as developed as its Cartesian counterpart. For this reason, we formulate here the fractional-order continuum theory assuming orthonormal curvilinear coordinates. The resulting approach allows leveraging the fractional-order constitutive relations \cite{patnaik2019FEM} in order to model the nonlocal interactions in curved structures. An example of the cylindrical shell system considered in the current study is illustrated in the schematic in Fig. \ref{fig: schematic}. While the assumption of orthonormal curvilinear coordinate system may seem somewhat restrictive, it does fit well the case of structures of practical interest in the aerospace field. 
Starting from the fractional-order continuum formulation developed for orthonormal Cartesian coordinates \cite{patnaik2019generalized}, we extend the formulation to curvilinear coordinates. This is followed by the development of a fractional-order shell theory based on the first-order shear deformation theory (FSDT) for displacement field variables \cite{reddy2006theory}. Finally, we undertake a numerical investigation to evaluate the effect of nonlocal interactions over the elastic response of cylindrical panels. 

\section{Fractional-order continuum theory in cylindrical coordinates}
\label{sec: constt_panel}
In this section, we develop the fractional-order continuum theory to describe the response of nonlocal solids in a cylindrical coordinate system. For this purpose, we extend the fractional-order kinematic approach \cite{patnaik2019generalized}, a sub-class of the displacement-driven approach to nonlocal elasticity \cite{patnaik2022displacement}, to a cylindrical coordinate basis. Recall that the key characteristic of the fractional-order kinematic approach consists in the fractional-order description of the strain-displacement relations which ultimately guarantees a positive-definite and well-posed approach. In the following, we first recall the definition of the strain and stress tensors following the fractional-order nonlocal continuum theory, and then proceed to cast these tensors in a cylindrical coordinate system. We refer the reader to \cite{patnaik2019generalized} for a detailed discussion on the formulation and physical interpretation of the fractional-order nonlocal continuum theory.

According to the fractional-order kinematic approach to nonlocal elasticity, the geometrically nonlinear strain tensor for the nonlocal solid is given by \cite{patnaik2019generalized,sidhardh2020geometrically}:
\begin{equation}
\label{eq: finite_fractional_strain}
\bm{\varepsilon} = \frac{1}{2} \left( \nabla^\alpha {\bm{u}} + \nabla^\alpha {\bm{u}}^{T} + \nabla^\alpha \bm{u} \nabla^\alpha \bm{u}^T \right)
\end{equation}
where $\bm{u}(\bm{x})$ denotes the displacement field. Further, $\nabla^\alpha(\cdot)$ denotes the fractional-order gradient operator which consists of Riesz-Caputo (RC) fractional-order derivatives, chosen specifically to describe the behavior of nonlocal solids \cite{patnaik2019generalized}. In fact, this definition guarantees frame-invariance at all points within the nonlocal continuum, including the material boundaries and interfaces; this important characteristic is not necessarily guaranteed by other definitions of the fractional-order operators. As evident from Eq.~(\ref{eq: finite_fractional_strain}), the application of the fractional-order nonlocal approach in cylindrical coordinates requires the definition of the fractional-order displacement gradient tensor using the cylindrical basis vectors.

\begin{figure}
    \centering
    \includegraphics[width=\textwidth]{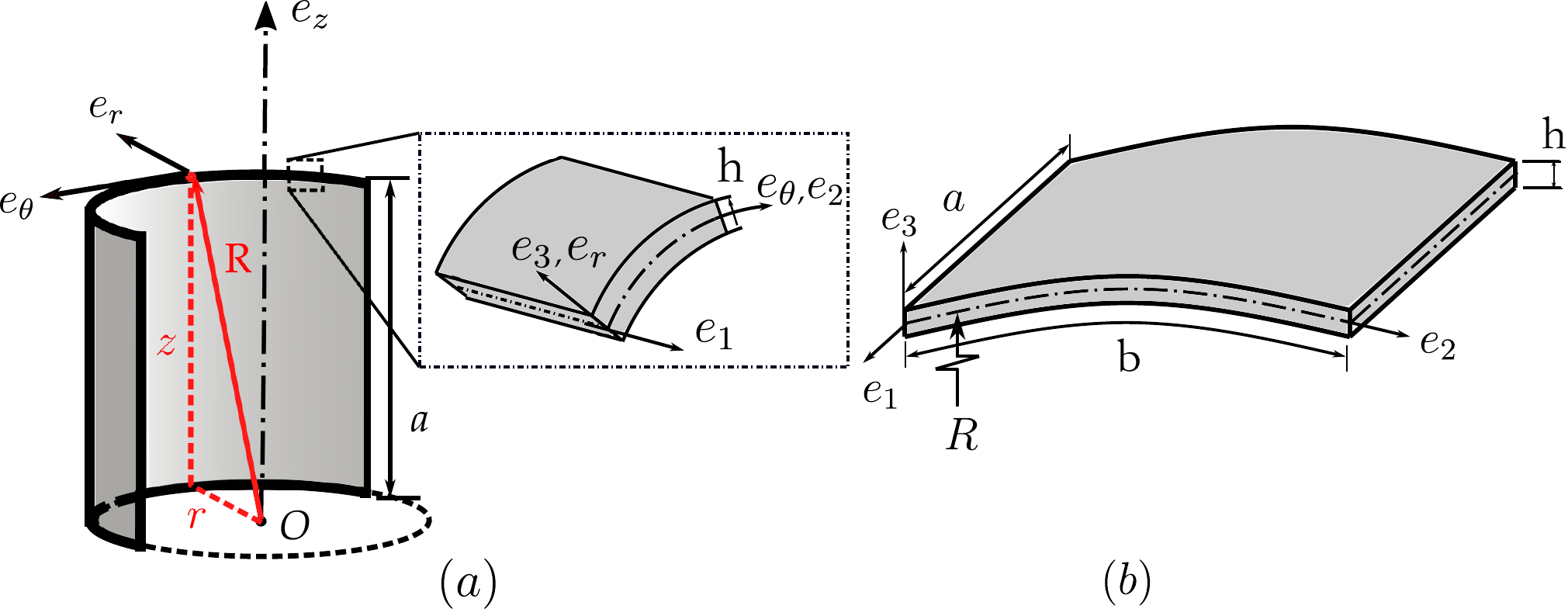}
    \caption{Coordinate axes (a) in cylindrical notation ($r-\theta-z$). Unit vectors in this system are denoted as $\hat{e}_r-\hat{e}_\theta-e_z$ (b) in curvilinear notation ($x_1-x_2-x_3$). Unit vectors in this coordinate system are depicted as $e_1-e_2-e_3$; depicted for the cylindrical shell panel of radius $R$ (not be confused with radius vector $\mathbf{R}$ indicated in red in Fig.~a). Note that while $\hat{e}_2$ and $\hat{e}_3$ in Fig.~b are identical to tangential $\hat{e}_{\theta}$ and normal (radial) $\hat{e}_{r}$ from Fig.~a, the axial coordinates $\hat{e}_{z}$ is replaced by $\hat{e}_{1}$ chosen to coincide with the mid-plane of the cylindrical shell. This transformation maintains the dimensional consistency for the deformation field components in \S~\ref{sec: nonlocal_shell}.}
    \label{fig: schematic}
\end{figure}

Consider the cylindrical coordinate system $r-\theta-z$ illustrated in Fig.~\ref{fig: schematic}a. The position vector of an arbitrary point in the cylindrical coordinate system is given as $\bm{R}=r\hat{e}_r+z\hat{e}_z$, where the basis vectors $\hat{e}_r$ and $\hat{e}_z$ denote the unit vectors along the radial and axial directions \cite{kreyszig1991introductory}. Note that the radial basis vector in the cylindrical coordinate system is a function of the azimuthal coordinate (here, denoted as $\theta$) such that $\hat{e}_r=\cos{\theta}\hat{i}+\sin{\theta}\hat{j}$. Similarly, the azimuthal basis vector in the cylindrical coordinate system, denoted here as $\hat{e}_\theta$, is also a function of $\theta$, that is $\hat{e}_\theta =-\sin{\theta}\hat{i}+\cos{\theta}\hat{j}$. The fractional exterior derivative of the position vector is given as \cite{tarasov2005fractional,cottrill2001fractional}:
\begin{equation}
\label{eq: ext_derivative_cylindrical}
    \mathrm{d}^\alpha \bm{R}=(\mathrm{d}r)^{\alpha} D^{\alpha}_{r} \bm{R}+(\mathrm{d}\theta)^{\alpha} D^{\alpha}_{\theta} \bm{R}+(\mathrm{d}z)^{\alpha} D^{\alpha}_{z} \bm{R}
\end{equation}
where $D^{\alpha}_{\xi}$ $(\xi=\{r,\theta,z\})$ is the RC fractional-order derivative and $\mathrm{d}\xi$ denotes the infinitesimal scalar increment in the $\xi$ direction. As mentioned above, the basis vectors $\hat{e}_r$ and $\hat{e}_\theta$ in the cylindrical coordinate system are not constant. In particular, the in-plane unit vectors ($\hat{e}_r$ and $\hat{e}_\theta$) are a function of the azimuthal coordinate $\theta$. The derivation of the fractional-order gradients in cylindrical coordinates require the fractional-order derivatives of these spatially-varying unit vectors with respect to the azimuthal variable ($\theta$), which are obtained as:
\begin{subequations}
\label{eq: ext_derivative_basis}
\begin{equation}
    D^{\alpha}_{\theta} \hat{e}_r = \underbrace{\left[ \mathcal{I}^{1-\alpha}_{\theta} \hat{e}_{\theta}\cdot \hat{e}_{r} \right]}_{\mathcal{F}_{r}}\hat{e}_r + \underbrace{\left[\mathcal{I}^{1-\alpha}_{\theta} \hat{e}_{\theta}\cdot \hat{e}_{\theta}\right]}_{\mathcal{F}_{\theta}} \hat{e}_\theta
\end{equation}
\begin{equation}
    D^{\alpha}_{\theta} \hat{e}_\theta = - \underbrace{\left[\mathcal{I}^{1-\alpha}_{\theta} \hat{e}_{\theta}\cdot \hat{e}_{\theta}\right]}_{\mathcal{F}_{\theta}} \hat{e}_r - \underbrace{\left[ \mathcal{I}^{1-\alpha}_{\theta} \hat{e}_{\theta}\cdot \hat{e}_{r} \right]}_{\mathcal{F}_{r}} \hat{e}_\theta
\end{equation}
\end{subequations}
where $\mathcal{F}_{r}$ and $\mathcal{F}_{\theta}$, as highlighted above, contain the scalar product of the radial and azimuthal unit vectors with the Reisz fractional integral $\mathcal{I}^{1-\alpha}_\theta \hat{e}_\theta$ (see supplementary derivations in Appendix). The detailed derivation of the fractional-order derivatives presented above is provided in the Appendix A. Note that for $\alpha=1$, corresponding to classical (local) elasticity, we obtain $\mathcal{F}_{r}=0$ and $\mathcal{F}_{\theta}=1$, such that the standard integer-order derivatives: $D^{1}_{\theta} \hat{e}_{r}=\hat{e}_{\theta}$ and $D^{1}_{\theta} \hat{e}_{\theta}=-\hat{e}_{r}$ are recovered\cite{kreyszig1991introductory}.
Utilizing the above results, the fractional exterior derivative of the position vector in Eq.~\eqref{eq: ext_derivative_cylindrical} is expressed in the following fashion:
\begin{equation}
    \label{eq: ext_derivative_cylindrical_simp}
    \mathrm{d}^\alpha \bm{R}=\left[(\mathrm{d}r)^{\alpha}+\mathcal{F}_{r} r (\mathrm{d}\theta)^{\alpha}\right]~\hat{e}_r +\left[\mathcal{F}_{\theta} r (\mathrm{d}\theta)^{\alpha}\right]~\hat{e}_\theta +\left[(\mathrm{d}z)^{\alpha}\right]~\hat{e}_z
\end{equation}

The exterior derivative of the displacement field $\bm{u}=u_r\hat{e}_r+u_\theta \hat{e}_\theta+u_z\hat{e}_z$ is given as follows:
\begin{equation}
\label{eq: ext_derivative_displacement}
    \mathrm{d}^{\alpha}\bm{u}=\mathrm{d}^{\alpha}u_r\hat{e}_r+\mathrm{d}^{\alpha}u_\theta \hat{e}_\theta+\mathrm{d}^{\alpha}u_z\hat{e}_z+u_r\mathrm{d}^{\alpha}\hat{e}_r+u_\theta \mathrm{d}^{\alpha}\hat{e}_\theta
\end{equation}
By using the definition for the exterior derivative in Eq.~\eqref{eq: ext_derivative_cylindrical} and the expressions for the fractional-order derivatives of the unit vectors in Eq. \eqref{eq: ext_derivative_basis}, the exterior derivative of the displacement vector can be further expressed as:
\begin{equation}
\begin{split}
    \mathrm{d}^{\alpha}\bm{u} = \left[ (\mathrm{d}r)^{\alpha} D^{\alpha}_{r} u_r + (\mathrm{d}\theta)^{\alpha} D^{\alpha}_{\theta} u_r +(\mathrm{d}z)^{\alpha} D^{\alpha}_{z} u_r \right] \hat{e}_r + \left[(\mathrm{d}r)^{\alpha} D^{\alpha}_{r} u_\theta + (\mathrm{d}\theta)^{\alpha} D^{\alpha}_{\theta} u_\theta +(\mathrm{d}z)^{\alpha} D^{\alpha}_{z} u_\theta \right] \hat{e}_\theta \\ + \left[ (\mathrm{d}r)^{\alpha} D^{\alpha}_{r} u_z + (\mathrm{d}\theta)^{\alpha} D^{\alpha}_{\theta} u_z +(\mathrm{d}z)^{\alpha} D^{\alpha}_{z} u_z \right] \hat{e}_r + u_r [(\mathrm{d}\theta)^\alpha D^{\alpha}_{\theta} \hat{e}_r] + u_\theta [(\mathrm{d}\theta)^\alpha D^{\alpha}_{\theta} \hat{e}_\theta]
\end{split}
\end{equation}
By using the expression for the fractional-order derivatives of the basis vectors in Eq.~\eqref{eq: ext_derivative_basis} the above expression is recast in the following manner:
\begin{equation}
    \label{eq: matrix_form_ext_der_disp}
    \underbrace{\begin{Bmatrix}
    (\mathrm{d}^{\alpha}\bm{u})\cdot \hat{e}_r\\
    (\mathrm{d}^{\alpha}\bm{u})\cdot \hat{e}_\theta\\
    (\mathrm{d}^{\alpha}\bm{u})\cdot \hat{e}_z\\
    \end{Bmatrix}}_{\text{Components of } \mathrm{d}^{\alpha}\bm{u}}
    =
    \begin{bmatrix}
    D^{\alpha}_{r} u_r & D^{\alpha}_{\theta} u_r+\mathcal{F}_{r} u_r-\mathcal{F}_{\theta} u_\theta & D^{\alpha}_{z} u_r\\
    D^{\alpha}_{r} u_\theta & D^{\alpha}_{\theta} u_\theta+\mathcal{F}_{\theta} u_r-\mathcal{F}_{r} u_\theta & D^{\alpha}_{z} u_\theta\\
    D^{\alpha}_{r} u_z & D^{\alpha}_{\theta} u_z & D^{\alpha}_{z} u_z\\
    \end{bmatrix}
    \begin{Bmatrix}
    (\mathrm{d}r)^{\alpha}\\
    (\mathrm{d}\theta)^{\alpha}\\
    (\mathrm{d}z)^{\alpha}\\
    \end{Bmatrix}
\end{equation}
The above expression will be used to derive the fractional-order displacement gradient tensor in the following.
Analogous to the classical integer-order vector calculus, the fractional-order gradient operator in the cylindrical coordinate system can be obtained from the following result: $\mathrm{d}^{\alpha}\bm{u}=\bm{\nabla}^{\alpha}\bm{u} \cdot \mathrm{d}^{\alpha}\bm{R}$. By using the definition for the external derivative of the position vector given in Eq.~\eqref{eq: ext_derivative_cylindrical_simp}, the external derivative of the displacement vector can be expressed as: 
\begin{equation}
    \label{eq: matrix_eqn_ext_derivative}
    \begin{Bmatrix}
    (\mathrm{d}^{\alpha}\bm{u})\cdot \hat{e}_r\\
    (\mathrm{d}^{\alpha}\bm{u})\cdot \hat{e}_\theta\\
    (\mathrm{d}^{\alpha}\bm{u})\cdot \hat{e}_z\\
    \end{Bmatrix}
    =
    \underbrace{\begin{bmatrix}
    D^{\alpha}_{r} u_r & D^{\alpha}_{\theta} u_r+\mathcal{F}_{r} u_r-\mathcal{F}_{\theta} u_\theta & D^{\alpha}_{z} u_r\\
    D^{\alpha}_{r} u_\theta & D^{\alpha}_{\theta} u_\theta+\mathcal{F}_{\theta} u_r-\mathcal{F}_{r} u_\theta & D^{\alpha}_{z} u_\theta\\
    D^{\alpha}_{r} u_z & D^{\alpha}_{\theta} u_z & D^{\alpha}_{z} u_z\\
    \end{bmatrix}
    \begin{bmatrix}
    1 & \frac{-\mathcal{F}_{r}}{\mathcal{F}_{\theta}} & 0\\
    0 & \frac{1}{\mathcal{F}_{\theta} r} & 0\\
    0 & 0 & 1
    \end{bmatrix}}_{\text{Components of tensor }\bm{\nabla}^{\alpha}\bm{u}}
    \underbrace{\begin{Bmatrix}
    (\mathrm{d}r)^{\alpha}+\mathcal{F}_{r} r (\mathrm{d}\theta)^{\alpha} \\
    \mathcal{F}_{\theta} r(\mathrm{d}\theta)^{\alpha}\\
    (\mathrm{d}z)^{\alpha}
    \end{Bmatrix}}_{\text{Components of } \mathrm{d}^{\alpha}\bm{R}}\\
\end{equation}
A direct comparison of the above expression with the result: $\mathrm{d}^{\alpha}\bm{u}=\bm{\nabla}^{\alpha}\bm{u} \cdot \mathrm{d}^{\alpha}\bm{R}$, provides the expression for the fractional-order displacement gradient. As highlighted in Eq.~\eqref{eq: matrix_eqn_ext_derivative}, the components of the fractional-order displacement gradient tensor are obtained from the product of the two matrices indicated above. By this, we obtain the fractional-order displacement gradient tensor in the cylindrical coordinate system as:
\begin{equation}
    \label{eq: grad_u_cylindrical}
    \begin{split}
    \bm{\nabla}^{\alpha}\bm{u}
    & = \left[ D^{\alpha}_{r} u_r \right] \hat{e}_r\otimes\hat{e}_r +  \left[-\frac{\mathcal{F}_{r}}{\mathcal{F}_{\theta}}D_r^{\alpha}u_r+\frac{1}{\mathcal{F}_{\theta}r}D_{\theta}^{\alpha}u_r+\frac{\mathcal{F}_{r}}{\mathcal{F}_{\theta}}\frac{u_r}{r}-\frac{u_\theta}{r}\right] \hat{e}_r\otimes\hat{e}_\theta + \left[D_{z}^{\alpha}u_r\right] \hat{e}_r\otimes\hat{e}_z \\
    & +\left[D^{\alpha}_{r} u_\theta\right] \hat{e}_\theta\otimes\hat{e}_r +  \left[-\frac{\mathcal{F}_{r}}{\mathcal{F}_{\theta}}D_r^{\alpha}u_\theta+\frac{1}{\mathcal{F}_{\theta}r}D_{\theta}^{\alpha}u_\theta+\frac{u_r}{r}-\frac{\mathcal{F}_{r}}{\mathcal{F}_{\theta}}\frac{u_\theta}{r}\right] \hat{e}_\theta \otimes\hat{e}_\theta + \left[ D^{\alpha}_{z} u_\theta \right] \hat{e}_\theta \otimes\hat{e}_z \\
    & +\left[D^{\alpha}_{r} u_z\right] \hat{e}_z\otimes\hat{e}_r + \left[-\frac{\mathcal{F}_{r}}{\mathcal{F}_{\theta}}D_r^{\alpha}u_z+\frac{1}{\mathcal{F}_{\theta}r}D_{\theta}^{\alpha}u_z\right]\hat{e}_z\otimes\hat{e}_\theta + \left[D^{\alpha}_{z} u_z\right]\hat{e}_z\otimes\hat{e}_z
    \end{split}
\end{equation}
where $'\otimes'$ denotes the dyadic tensor product. Note that the fractional-order gradient operator derived here is highly general in nature and can be used to evaluate the fractional gradient of any scalar or vector field, and not necessarily the displacement field. 

The above derivation of the fractional-order displacement gradient tensor in the cylindrical coordinate system enables a direct implementation of the fractional-order nonlocal approach for the analysis of nonlocal structures better represented or constructed in the cylindrical frame of reference. It remains to define the stress, corresponding to the nonlocal strain, in the cylindrical coordinate system. According to the fractional-order kinematic approach to nonlocal elasticity, the stress in the nonlocal solid is defined as:
\begin{equation}
\label{eq: stress}
    \bm{\sigma} = \mathbb{C}:\bm{\varepsilon}
\end{equation}
where $\bm{\sigma}$ and $\bm{\varepsilon}$ are the nonlocal stress and fractional-order strain expressed in cylindrical coordinates, and $\mathbb{C}$ denotes the fourth-order material elasticity tensor The nonlocal strain in the above expression follows from the substitution of the result given in Eq.~\eqref{eq: grad_u_cylindrical} in Eq.~\eqref{eq: finite_fractional_strain}. Note that the stress defined through the above expression is intrinsically nonlocal due to the differ-integral nature of the fractional-order gradient operators within the definition of the nonlocal strain. In other terms, the effect of the nonlocal interactions from the horizon of nonlocality at a given point is accounted within the strain tensor via the fractional-order operators. In addition to the established advantages associated with the use of a fractional order kinematic approach \cite{patnaik2019generalized,patnaik2022displacement}, this method also ensures a direct application of the stress-strain constitutive relationship (Eq.~(\ref{eq: stress})). The result is a direct evaluation of the stress field from the strain field, irrespective of the nature of the coordinate system. This is in sharp contrast to classical integral approaches where the integral nature of the stress-strain constitutive relationship warrants additional mathematical transformations (or, manipulations) for the evaluation of the stress field from the strain field. This characteristic presents significant challenges for modeling nonlocal cylindrical panels via classical integral approaches, also highlighted by the lack of studies in this area.

\section{Fractional-order shell theory of nonlocal cylindrical panels}
\label{sec: nonlocal_shell}
In this section, we use the fractional-order continuum theory in cylindrical coordinates, developed in the previous section, to formulate the fractional-order shell theory of nonlocal cylindrical panels. A thorough review of the literature indicates that the curvilinear coordinate system is widely adopted in modeling the response of local as well as nonlocal cylindrical panels. Indeed the uniformity of dimensionality (that is, the dimension of length) across the bases of the curvilinear coordinate system allows for a simpler theoretical analysis and numerical simulation when compared to the cylindrical coordinate system. Hence, in this study, we develop the fractional-order shell theory in the curvilinear coordinate system.

The schematic of the cylindrical panel within the curvilinear coordinate axes is provided in Fig.~\ref{fig: schematic}b. The length, width, thickness and radius of the panel are denoted by $a$, $b$, $h$, and $R$ \footnote{Note the difference in notation of the radius of the panel $R$ and the position vector $\bm{R}$.}, respectively. As illustrated in the schematic, the mid-plane of the panel is considered as the reference plane, that is, $x_3=0$. Consequently, the top and bottom surface of the panel are identified as $x_{3}=\pm h/2$. The origin of the curvilinear coordinate system is chosen such that $x_1=0$ and $x_1=a$ indicate the straight edges of the panel, and $x_2=0$ and $x_2=b$ indicate the curved edges. According to the first-order shear deformation theory, the displacement field at any spatial location $\bm{x}(x_1,x_2,x_3)$ on the panel is related to the displacement field of the mid-plane of the panel in the following manner \cite{reddy2006theory}:
\begin{subequations}
\label{eq: disp_field}
\begin{equation}
    u_1(\bm{x})=u_0(\bm{x}_0)+x_3\theta_0(\bm{x}_0)
\end{equation}
\begin{equation}
    u_2(\bm{x})=v_0(\bm{x}_0)+x_3\phi_0(\bm{x}_0)
\end{equation}
\begin{equation}
\label{eq: tranverse_displacment}
    u_3(\bm{x})=w_0(\bm{x}_0)
\end{equation}
\end{subequations}
In the above equation, $u_i~(i=1,2,3)$ are the components of displacement field $\bm{u}$ expressed in the curvilinear coordinate system (see Fig.~\ref{fig: schematic}b). Further, $u_0$, $v_0$, $w_0$, $\theta_0$ and $\phi_0$ denote the generalized displacement variables corresponding to the translation and rotation at a point $\bm{x}_0(x_1,x_2)$ located on the mid-plane $x_3=0$. Hereafter, for a compact notation, we do not specify the functional dependence of the displacement coordinates on $\bm{x}_0(x_1,x_2)$. 

In order to derive the expression for the strain field, corresponding to the displacement field in Eq.~(\ref{eq: disp_field}), we transform the expression for the fractional-order displacement gradient tensor in Eq.~(\ref{eq: grad_u_cylindrical}) from the cylindrical coordinate system to the curvilinear coordinate system. This change in the basis involves the use of a set of three transformations $\{r\to R+x_3,~\theta\to x_2/R,~z\to x_1\}$. The transformed fractional-order displacement gradient tensor, expressed in the curvilinear coordinates, is obtained as:
\begin{equation}
    \label{eq: grad_u_curv}
    \begin{split}
    \bm{\nabla}^{\alpha}\bm{u} & =
    \left[D_{x_1}^{\alpha}u_1\right] \hat{e}_{1}\otimes\hat{e}_{1} + \left[-\frac{\mathcal{F}_{r}}{\mathcal{F}_{\theta}}D_{x_3}^{\alpha}u_1+\frac{1}{\mathcal{F}_{\theta}}D_{x_2}^{\alpha}u_1\right] \hat{e}_{1}\otimes\hat{e}_{2} + \left[D_{x_3}^{\alpha}u_1\right] \hat{e}_{1}\otimes\hat{e}_{3}\\
    & + \left[ D_{x_1}^{\alpha}u_2 \right]\hat{e}_{2}\otimes\hat{e}_{1} + \left[-\frac{\mathcal{F}_{r}}{\mathcal{F}_{\theta}}D_{x_3}^{\alpha}u_2+\frac{1}{\mathcal{F}_{\theta}}D_{x_2}^{\alpha}u_2+\frac{u_3}{R}-\frac{\mathcal{F}_{r}}{\mathcal{F}_{\theta}}\frac{u_2}{R}\right] \hat{e}_{2}\otimes\hat{e}_{2} + \left[D_{x_3}^{\alpha}u_2\right]\hat{e}_{2}\otimes\hat{e}_{3} \\
    & + \left[D_{x_1}^{\alpha}u_3\right]\hat{e}_{3}\otimes\hat{e}_{1} + \left[-\frac{\mathcal{F}_{r}}{\mathcal{F}_{\theta}}D_{x_3}^{\alpha}u_3+\frac{1}{\mathcal{F}_{\theta}}D_{x_2}^{\alpha}u_3+\frac{\mathcal{F}_{r}}{\mathcal{F}_{\theta}}\frac{u_3}{R}-\frac{u_2}{R}\right] \hat{e}_{3}\otimes\hat{e}_{2} + \left[D_{x_3}^{\alpha}u_3\right] \hat{e}_{3}\otimes\hat{e}_{3}
    \end{split}
\end{equation}
The derivation of the fractional-order displacement gradient tensor in the curvilinear coordinate system from the cylindrical coordinate system follows standard principles of coordinate transformation and can be found readily in literature \cite{amabili2008nonlinear}. In order to avoid unnecessary repetitions, we do not report the detailed steps here.
In deriving the above expression, we assumed that the panel is shallow in nature, that is, $h \ll R$. This assumption is consistent with the geometry seen in practical applications for curved structures and allows higher-order powers of $x_3/R$ and $h/R$ to be neglected in the subsequent formulation. $D^\alpha_{x_i}(\cdot)$ in Eq.~(\ref{eq: grad_u_curv}) denotes the RC fractional derivative of order $\alpha$. The RC derivative, at a given point ${\bm{x}}$, is defined on the interval $[{x}_i-l_{-_i},{x}_i+l_{+_i}] \equiv [x_{-_i},x_{+_i}]$ in the $\hat{e}_i$ direction as:
\begin{equation}
\label{eq: RC_definition}
	D^{\alpha}_{x_i} u_j({{\bm{x}}})=\frac{1}{2}\Gamma(2-\alpha) \left[ l_{-_i}^{\alpha-1} \left( {}^C_{x_{-_i}}D^{\alpha}_{x_i} u_j({{\bm{x}}},t) \right) - l_{+_i}^{\alpha-1} \left( {}^C_{x_i}D^{\alpha}_{x_{+_i}} {u_j}({\bm{x}},t) \right) \right]
\end{equation}
where $\Gamma(\cdot)$ is the Gamma function. ${}^C_{x_{-_i}}D^{\alpha}_{x_i} u_j$ and ${}^C_{x_j}D^{\alpha}_{x_{+_j}} u_i$ are the left- and right-handed fractional-order Caputo derivatives of $u_i$, respectively. Further, $l_{-_i}$ and $l_{-_i}$ denote the length scales of the fractional-order continuum formulation and characterize the horizon of nonlocality of the cylindrical panel \cite{patnaik2019generalized}.

We will derive the displacement gradient relations for a shear-deformable nonlocal cylindrical shell by using the displacement field in Eq.~(\ref{eq: disp_field}) within Eq.~(\ref{eq: grad_u_curv}). Assuming moderate rotations of the transverse normals $(10^\circ-15^\circ)$ and small displacement gradients, we obtain the nonlinear von-K\'arm\'an strain-displacement relations for the cylindrical panel from the nonlocal strain tensor definition in Eq.~(\ref{eq: finite_fractional_strain}) as:
\begin{subequations}
\label{eq: frac_strains_xyz}
\begin{equation}
    \varepsilon_{11}(\bm{x})=D_{x_1}^{\alpha}u_0+x_3D_{x_1}^{\alpha}\theta_0 + 
    \underbrace{\frac{1}{2}\left(D_{x_1}w_0\right)^2}_{\text{Nonlinear}}
\end{equation}
\begin{equation}
\begin{split}
    \varepsilon_{22}(\bm{x})=&\frac{1}{\mathcal{F}_{\theta}}\left(D_{x_2}^{\alpha}v_0+x_3D_{x_2}^{\alpha}\phi_0\right)+\frac{w_0}{R}-\frac{\mathcal{F}_{r}}{\mathcal{F}_{\theta}}\left[\phi_0 + \left( \frac{v_0+x_3\phi_0}{R} \right) \right]\\
    &+\underbrace{\frac{1}{2}\left[-\frac{\mathcal{F}_{r}}{\mathcal{F}_{\theta}}D_{x_3}^{\alpha}w_0+\frac{1}{\mathcal{F}_{\theta}}D_{x_2}^{\alpha}w_0+\frac{\mathcal{F}_{r}}{\mathcal{F}_{\theta}}\frac{w_0}{R} - \left( \frac{v_0+x_3\phi_0}{R} \right) \right]^2}_{\text{Nonlinear}}
\end{split}
\end{equation}
\begin{equation}
\begin{split}
    {\gamma}_{12}(\bm{x})=2\varepsilon_{12}(\bm{x})=&D_{x_1}^{\alpha}v_0+x_3D_{x_1}^{\alpha}\phi_0+\left[-\frac{\mathcal{F}_{r}}{\mathcal{F}_{\theta}}\theta_0+\frac{1}{\mathcal{F}_{\theta}}\left(D_{x_2}^{\alpha}u_0+x_3D_{x_2}^{\alpha}\theta_0\right)\right] \\
    &+\underbrace{\left[-\frac{\mathcal{F}_{r}}{\mathcal{F}_{\theta}}D_{x_3}^{\alpha}w_0+\frac{1}{\mathcal{F}_{\theta}}D_{x_2}^{\alpha}w_0+\frac{\mathcal{F}_{r}}{\mathcal{F}_{\theta}}\frac{w_0}{R} - \left( \frac{v_0+x_3\phi_0}{R} \right) \right] D_{x_1}^{\alpha}w_0}_{\text{Nonlinear}}
\end{split}
\end{equation}
\begin{equation}
    {\gamma}_{13}(\bm{x})=2\varepsilon_{13}(\bm{x})=D_{x_1}^{\alpha}w_0+\theta_0
\end{equation}
\begin{equation}
    {\gamma}_{23}(\bm{x})=2\varepsilon_{23}(\bm{x})=\phi_0+ \left[\frac{D_{x_2}^{\alpha}w_0}{\mathcal{F}_{\theta}}+\frac{\mathcal{F}_{r}}{\mathcal{F}_{\theta}}\frac{w_0}{R}- \left( \frac{v_0+x_3\phi_0}{R} \right) \right]
\end{equation}
\end{subequations}
It follows from Eq.~(\ref{eq: tranverse_displacment}) that the transverse normal strain $\varepsilon_{33}(\bm{x})$ is identically zero, which is consistent with the assumptions of the first-order shear deformation theory \cite{reddy2006theory}. Further, observe that the transverse shear strains, unlike the in-plane normal and shear strains, do not show a nonlocal dependence on the rotations of the mid-plane of the panel. This is a direct result of the uniform variation of the displacement degrees of freedom ($\{u_0,v_0,w_0,\theta_0,\phi_0\}$) across the thickness of the panel when adopting the first-order shear deformation theory (see Eq.~(\ref{eq: disp_field})). From a physical perspective, the effect of the nonlocal interactions across the thickness of the slender panel are negligible when compared to the nonlocal interactions across its plane. 

The stress field, corresponding to the strain field in Eq.~(\ref{eq: frac_strains_xyz}), is determined using the linear stress-strain relationship given in Eq.~(\ref{eq: stress}). Using the strain and stress fields, we obtain the nonlocal strain energy $\mathbb{U}$ and the work done by the externally applied forces $\mathbb{V}$:
\begin{subequations}
\label{eq: Virtual_quantities}
\begin{equation}
\label{eq: Virtual_strain_energy}
    \mathbb{U} = \frac{1}{2}\int_\Omega \int_{-\frac{h}{2}}^{\frac{h}{2}} \left[ \sigma_{11}\varepsilon_{11} + \sigma_{22}\varepsilon_{22} + \sigma_{12}\gamma_{12} + \sigma_{13}\gamma_{13} + \sigma_{23}\gamma_{23} \right]\mathrm{d}x_3 \mathrm{d} \Omega
\end{equation}
\begin{equation}
\label{eq: Virtual_work}
    \mathbb{V} = \int_\Omega \left[ F_{x_1} u_0 + F_{x_2} v_0 + F_{x_3} w_0 + M_{x_1} \theta_0 + M_{x_2} \phi_0 \right]\mathrm{d} \Omega
\end{equation}
\end{subequations}
where $\mathrm{d}\Omega$ denotes an infinitesimal element on the mid-plane of the panel. $\{F_{x_1},F_{x_2},F_{x_3}\}$ are the external loads applied per unit area of the mid-plane of the panel in the $\hat{e}_1$, $\hat{e}_2$ and $\hat{e}_3$ directions, respectively. $\{M_{x_1},M_{x_2}\}$ are the external moments applied per unit area. By minimizing the total potential energy of the panel (that is, $\Pi=\mathbb{U}-\mathbb{V}$) using variational principles, the nonlinear fractional-order governing equations and the boundary conditions of the nonlocal cylindrical panel are obtained as:
\begin{subequations}
\label{eq: GDE}
\begin{equation}
\mathfrak{D}^\alpha_{x_1} N_{11} + \mathfrak{D}^\alpha_{x_2} \left(\frac{N_{12}}{\mathcal{F}_\theta}\right) + F_{x_1} = 0
\end{equation}
\begin{equation}
\mathfrak{D}^\alpha_{x_1} N_{12} + \mathfrak{D}^\alpha_{x_2} \left(\frac{N_{22}}{\mathcal{F}_\theta}\right)+\frac{Q_{23}}{R} + F_{x_2} = 0
\end{equation}
\begin{equation}
\mathfrak{D}^\alpha_{x_1} \left[ Q_{13} + N_{11} D^\alpha_{x_1} w_0 + \left(\frac{N_{12}}{\mathcal{F}_\theta}\right) D^\alpha_{x_2} w_0 \right] + 
\mathfrak{D}^\alpha_{x_2} \left[ \frac{Q_{23}}{\mathcal{F}_\theta} +  \left(\frac{N_{12}}{\mathcal{F}_\theta}\right) D^\alpha_{x_1} w_0 + \left(\frac{N_{22}}{\mathcal{F}^2_\theta}\right) D^\alpha_{x_2} w_0 \right]-\frac{N_{22}}{R} +
F_{x_3} = 0
\end{equation}
\begin{equation}
\mathfrak{D}^\alpha_{x_1} M_{11} + \mathfrak{D}^\alpha_{x_2} \left(\frac{M_{12}}{\mathcal{F}_\theta}\right) - Q_{13} + M_{x_1} = 0
\end{equation}
\begin{equation}
\mathfrak{D}^\alpha_{x_1} M_{12} + \mathfrak{D}^\alpha_{x_2} \left(\frac{M_{22}}{\mathcal{F}_\theta}\right) - Q_{23} + M_{x_2} = 0
\end{equation}
\end{subequations}
where $\{N_{11}, N_{22}, N_{12}\}$ are the in-plane stress resultants, $\{Q_{13},Q_{23}\}$ are the transverse shear stress resultants, and $\{M_{11},M_{22},M_{12}\}$ are the moment resultants. The corresponding essential and natural boundary conditions are obtained as:
\begin{subequations}
\label{eq: BC}
\begin{equation}
\forall \bm{x} \in \delta\Omega_{x_1} \left\{ \begin{matrix}
\mathcal{I}_{x_1}^{1-\alpha}N_{11}=0~\text{or}~\delta u_0=0,~~~\mathcal{I}_{x_1}^{1-\alpha}N_{12}=0~\text{or}~\delta v_0=0,~~~\mathcal{I}_{x_1}^{1-\alpha}(Q_{13} + \mathcal{N}_{11})=0~\text{or}~\delta w_0=0
\\
\mathcal{I}_{x_1}^{1-\alpha}M_{11}=0~\text{or}~\delta \theta_0=0,~~~\mathcal{I}_{x_1}^{1-\alpha}M_{12}=0~\text{or}~\delta \phi_0=0
\end{matrix}\right.   
\end{equation}
\begin{equation}
\forall \bm{x} \in \delta\Omega_{x_2}  \left\{ \begin{matrix}
\mathcal{I}_{x_2}^{1-\alpha}\left(\frac{N_{12}}{\mathcal{F}_\theta}\right)=0~\text{or}~\delta u_0=0,~~~\mathcal{I}_{x_2}^{1-\alpha}\left(\frac{N_{22}}{\mathcal{F}_\theta}\right)=0~\text{or}~\delta v_0=0,~~~\mathcal{I}_{x_2}^{1-\alpha}\left(\frac{Q_{23}}{\mathcal{F}_\theta} + \mathcal{N}_{22}\right)=0~\text{or}~\delta w_0=0
\\
\mathcal{I}_{x_2}^{1-\alpha}\left(\frac{M_{12}}{\mathcal{F}_\theta}\right)=0~\text{or}~\delta \theta_0=0,~~~\mathcal{I}_{x_2}^{1-\alpha}\left(\frac{M_{22}}{\mathcal{F}_\theta}\right)=0~\text{or}~\delta \phi_0=0
\end{matrix}
\right.   
\end{equation}
\end{subequations}
where $\mathcal{N}_{11} = N_{11} D^\alpha_{x_1} w_0 + (N_{12}/\mathcal{F}_\theta) D^\alpha_{x_2} w_0$ and $\mathcal{N}_{22} = (N_{12}/\mathcal{F}_\theta) D^\alpha_{x_1} w_0 + (N_{22}/\mathcal{F}^2_\theta) D^\alpha_{x_2} w_0$. In the above derivation, we have employed the shallow shell assumption ($(x_3/R)$ and $u_i/R$ are $\ll 1$). Also, numerical evaluation of the parameter $\mathcal{F}_r$ following the relation in the Appendix presents $\mathcal{F}_r\ll1$ $\forall \alpha,~l_\theta$. So, the terms with this parameter in the above equations can be ignored without loss of accuracy.

The different stress and moment resultants in the above governing equations are defined analogously to the classical (local) formulation as:
\begin{subequations}
\label{eq: stress_resultants}
\begin{equation}
    \{ N_{11}, N_{22}, N_{12}, Q_{13}, Q_{23}\} = \int_{-\frac{h}{2}}^{\frac{h}{2}} \{ \sigma_{11}, \sigma_{22}, \sigma_{12}, K_s \sigma_{13}, K_s\sigma_{23}\} \mathrm{d}z
\end{equation}
\begin{equation}
    \{M_{11}, M_{22}, M_{12}\} = \int_{-\frac{h}{2}}^{\frac{h}{2}} \{ z \sigma_{11}, z \sigma_{22}, z\sigma_{12} \}\mathrm{d}z
\end{equation}
\end{subequations}
where $K_s$ is the shear correction factor. The shear correction factor is chosen as $K_s=5/6$, equal to the value adopted in classical elasticity theory \cite{reddy2006theory}. Recently, it was shown that the shear correction factor for nonlocal solids, particularly heterogeneous multiscale solids with functional gradation in either the material or the geometric properties, varies from $K_s=5/6$ following a non-classical redistribution of stress through the thickness as a result of nonlocality \cite{ding2021multiscale}. However, we note that the difference in the shear correction factor for nonlocal solids from the classical value $K_s=5/6$ is typically around $1\%$, depending on the specific functional variation of the material properties. We emphasize that this difference does not (drastically) alter the numerical predictions as well as the general conclusions (on the effect of nonlocality on response of shells), presented in this study.

Finally, the Riesz fractional integral $\mathcal{I}^{1-\alpha}_{x_i}(\cdot)$ in the governing equations is defined as:
\begin{equation}
\label{eq: reisz integral_def}
    \mathcal{I}^{1-\alpha}_{x_i} \psi =\frac{1}{2}\Gamma(2-\alpha) \left[ l_{+_i}^{\alpha-1} \left( {}_{x_i-l_{+_i}}\mathcal{I}_{x_i}^{1-\alpha} \psi \right) - l_{-_i}^{\alpha-1} \left( {}_{x_i}\mathcal{I}_{x_i+l_{-_i}}^{1-\alpha} \psi \right) \right]
\end{equation}
where ${}_{x_i-l_{+_i}}\mathcal{I}_{x_i}^{1-\alpha}\psi$ and ${}_{x_i}\mathcal{I}_{x_i+l_{-_i}}^{1-\alpha}\psi$ are the left- and right-handed Riesz integrals to the order $\alpha$ of an arbitrary function $\psi$, respectively. The fractional-order derivative $\mathfrak{D}_{x_i}^\alpha(\cdot)$ is the first-order derivative of the Riesz integral:
\begin{equation}
    \mathfrak{D}^\alpha_{x_i} \psi = D^1_{x_i} \left[ \mathcal{I}^{1-\alpha}_{x_i} \psi \right]
\end{equation}

The detailed derivation of the nonlinear fractional-order governing equations via variational principles can be found in \cite{sidhardh2020geometrically,patnaik2020geometrically} for nonlocal beams and plates. We emphasize that the same variational techniques are directly applicable here to derive the strong-form governing equations for the nonlocal cylindrical panel. Note that the strong-form of the governing equations of the shell and of the corresponding boundary conditions can also be expressed in terms of the displacement field variables by using the constitutive stress-strain relations of the plate along with the stress and moment resultants given in Eq.~(\ref{eq: stress_resultants}). This procedure is routine (see, for example, \cite{patnaik2020geometrically}) and hence, we do not provide these details here. Finally, as expected, the classical geometrically nonlinear shell governing equations and the corresponding boundary conditions are recovered for $\alpha=1$. 

\section{Results and discussion}
\label{sec: results}
In this section, we use the previously developed formulation to investigate numerically the response of nonlocal shells and to provide a quantitative assessment of the influence of the nonlocal interactions on the linear and geometrically nonlinear response of cylindrical panels. We primarily focus on the effect of the constitutive parameters corresponding to fractional-order constitutive theory: the fractional-order and horizon of nonlocality. Moreover, the effect of the coupling between the nonlocal interactions and the curvature of the panels is also investigated. 

\begin{figure}[hb]
    \centering
    \includegraphics[width=0.6\textwidth]{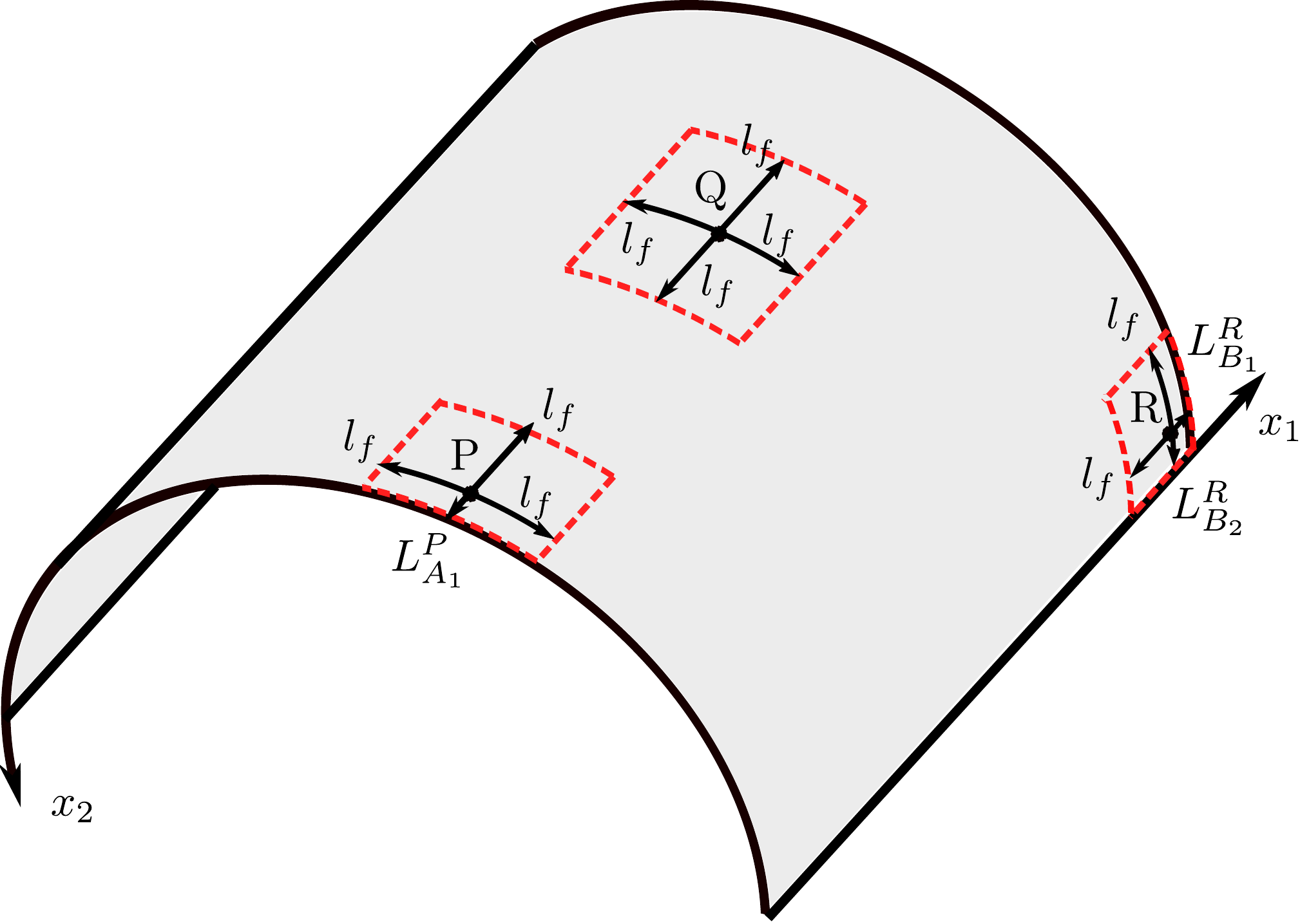}
    \caption{Position-dependent nature of the horizon of nonlocal influence illustrated for three points ($P$, $Q$ and $R$) in an isotropic shell. Note the asymmetry in the horizon length when points are close to the boundaries (e.g. $P$ and $R$).}
    \label{fig: length_trunc} 
\end{figure}

For all the simulations conducted in this study, we assume that the cylindrical panel is made out of an isotropic material with Young's modulus $E=30$ MPa and Poisson's ratio $\nu=0.3$. The length and width of the panel are assumed to be $a=b=1$m, and the thickness is assumed to be $h=a/10$. Unless specifically mentioned, we assume the radius of curvature to be $R=10a$. This choice of the geometric parameters follows from the assumption of shallow shells ($h/R=0.01\ll 1$) in Eq.~\eqref{eq: grad_u_curv}. We emphasize that the choice of both the material and geometric parameters is arbitrary, while considering shallow cylindrical shells, and the framework can handle any general choice of material properties. The fractional-order and the nonlocal length scales are not fixed \textit{a priori}, and will be treated as parameters to determine their effect on the structural response. Consequently, their numerical values will be specified wherever necessary. Note that the nonlocal length scales for a general point within the cylindrical panel follow: $l_{-_1}=l_{+_1}=l_{1}$ and $l_{-_2}=l_{+_2}=l_{2}$. The symmetry of this parameter (in $x_1$ and $x_2$-directions) around a given point is broken for points close to the boundaries. In such cases, the nonlocal horizon of influence is truncated at the external boundary encountered in order to achieve a physically consistent and frame invariant formulation \cite{patnaik2019generalized}. This latter aspect is illustrated schematically in Fig.~\ref{fig: length_trunc}. Additionally, we also assume an isotropic horizon of nonlocality, such that $l_{\square}=l_{f},~\square\in\{1,2\}$.

As mentioned previously, we will numerically investigate both the linear and the geometrically nonlinear response of nonlocal cylindrical panels. In each study, the panels are subject to a uniformly distributed transverse load (UDTL) applied on the top surface of the panel. The magnitude of the UDTL is varied, and its value will be provided whenever required. Unless otherwise mentioned, the loads are applied along $+\hat{e}_3$-direction (see Fig.~\ref{fig: schematic}). We analyze the effect of two different boundary conditions on the response of the nonlocal panel. First, we consider a fully clamped (CCCC) panel where all the transverse edges are subjected to the following constraints on the generalized displacement coordinates (defined in Eq.~\eqref{eq: disp_field}) \cite{reddy2006theory}:
\begin{equation}
\label{eq: CCCC}
\begin{matrix} \forall x_1=\{0,a\}: & u_0=v_0=w_0=\theta_0=\phi_0=0 \\ \forall x_2=\{0,b\}: & u_0=v_0=w_0=\theta_0=\phi_0=0 \end{matrix}
\end{equation}
Next, we consider a simply supported (SSSS) panel with the following constraints on the generalized displacement coordinates \cite{reddy2006theory}:
\begin{equation}
\label{eq: SSSS}
\begin{matrix} \forall x_1=\{0,a\}: & v_0=w_0=\phi_0=0 \\ \forall x_2=\{0,b\}: & u_0=w_0=\theta_0=0 \end{matrix}
\end{equation}

\begin{figure*}[b!]
    \centering
    \begin{subfigure}[t]{0.48\textwidth}
        \centering
        \includegraphics[width=\textwidth]{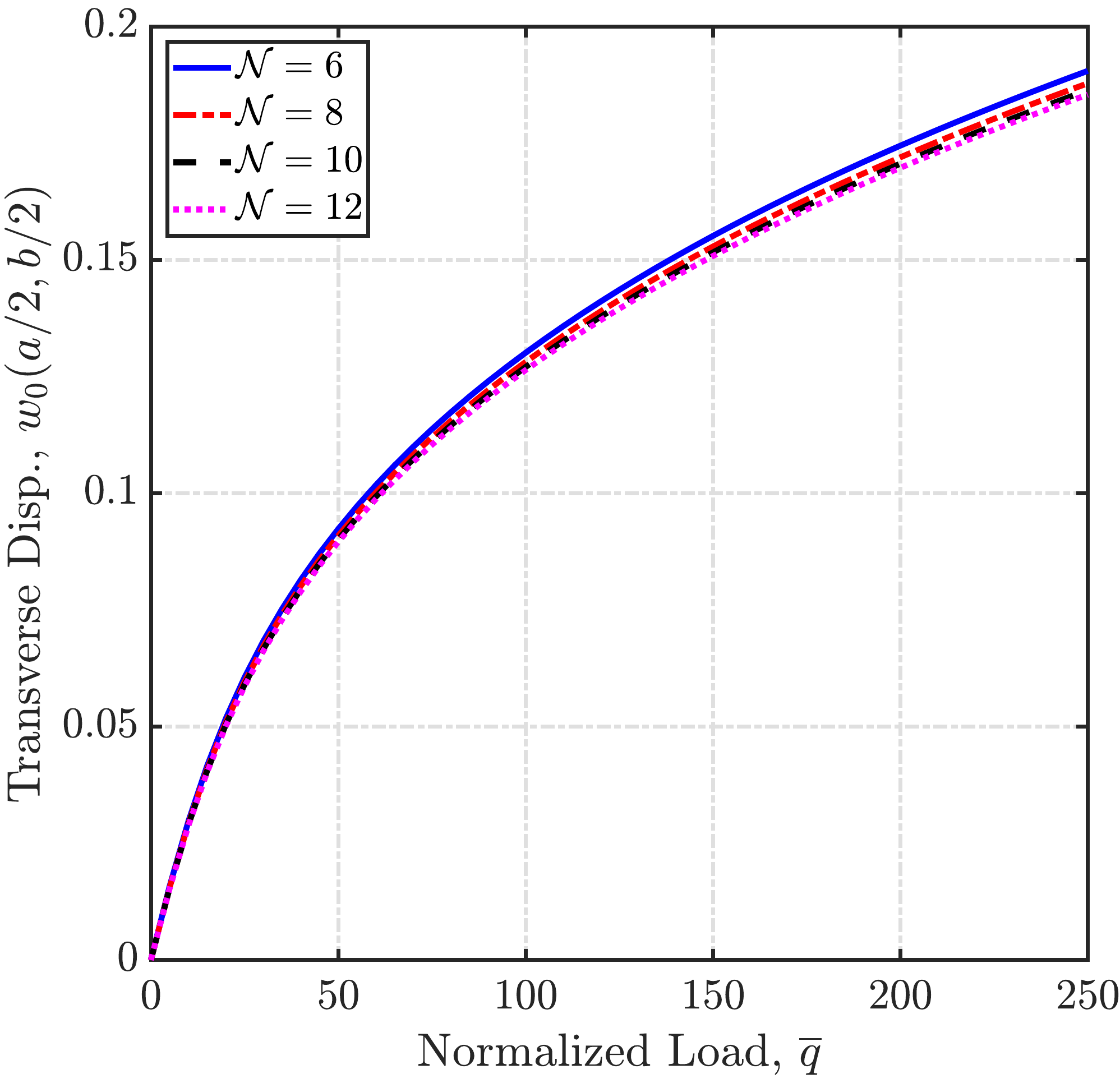}
        \caption{$\alpha=0.8$, $l_f/a=0.5$.}
        \label{fig: conv_1}
    \end{subfigure}
    ~
    \begin{subfigure}[t]{0.48\textwidth}
    \centering
        \includegraphics[width=\textwidth]{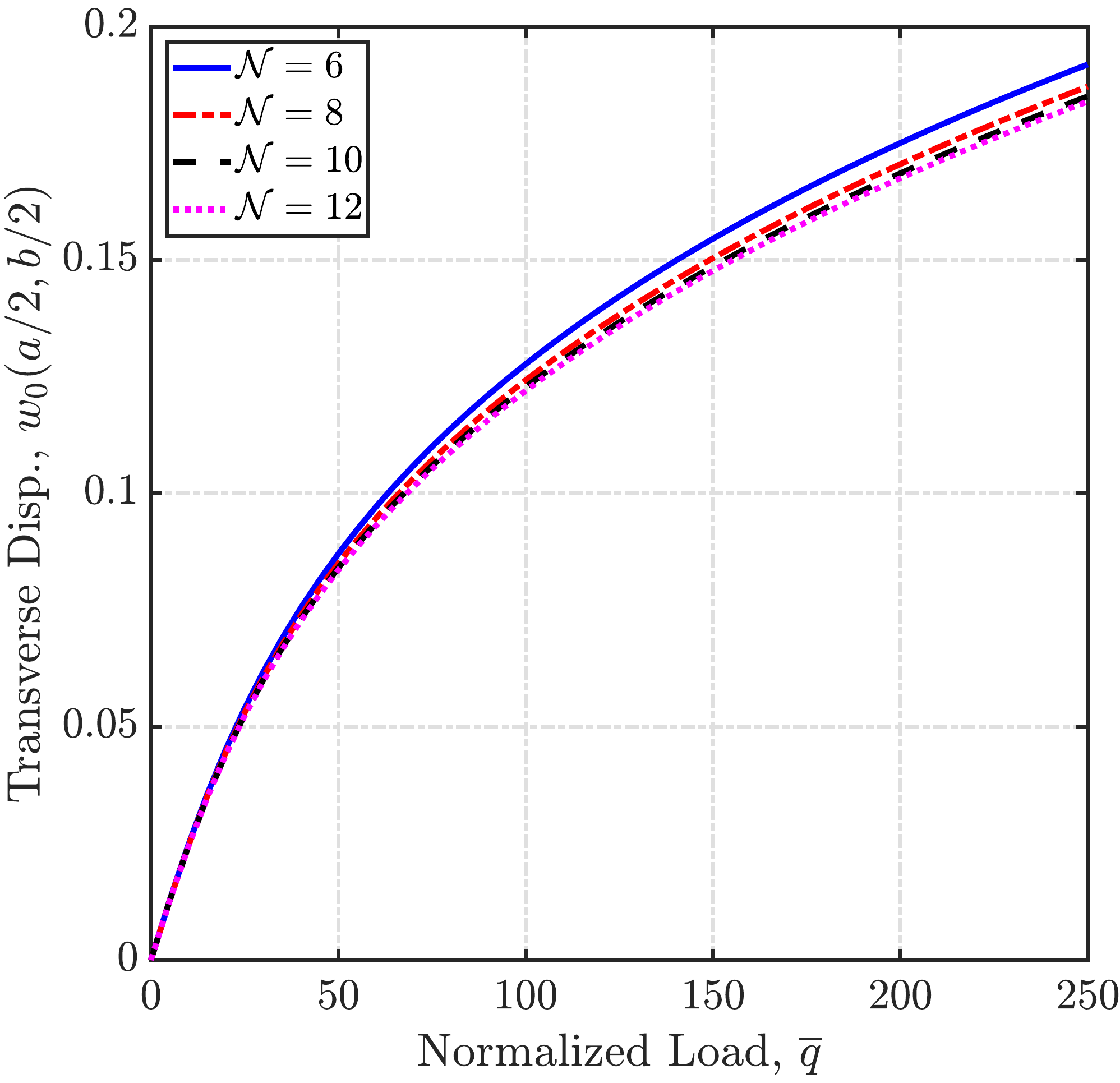}
        \caption{$\alpha=0.9$, $l_f/a=1.0$.}
        \label{fig: conv_2}
    \end{subfigure}%
    \caption{Load-displacement curves for the geometrically nonlinear response of clamped cylindrical panels of radius $R/a=10$. Convergence of the numerical code is established with $<1\%$ difference in results upon increasing discretization of the finite elements beyond $\mathcal{N}_1=\mathcal{N}_2=10$. Note that the transverse load is non-dimensionalized following Eq.~\eqref{eq: load_nondim}.}
    \label{fig: convergence}
\end{figure*}
In order to numerically simulate the fractional-order nonlinear governing equations of the nonlocal cylindrical panel, we leverage the fractional-order finite element method (f-FEM) developed in \cite{patnaik2019FEM}. Analogous to classical finite element techniques, the f-FEM converts the nonlinear fractional-order partial differential equations into a set of nonlinear algebraic equations which are solved using an incremental Newton-Raphson method. The detailed algorithm of the f-FEM and the iterative Newton-Raphson method can be found in detail in \cite{sidhardh2020geometrically, patnaik2020geometrically}. For the sake of brevity, we provide only a brief discussion of the f-FEM for the simulation of nonlocal shells in Appendix B.

Before proceeding further, we conduct a convergence study to ascertain the appropriate choice for mesh discretization. For the convergence study, we restrict ourselves to the geometrically nonlinear response of the panel. The load-displacement curves corresponding to two arbitrarily chosen values of the nonlocal constitutive parameters: (i) $\alpha=0.8$ and $l_f/a=0.5$, and  (ii) $\alpha=0.9$ and $l_f/a=1.0$, are presented in Fig.~\ref{fig: conv_1} and Fig.~\ref{fig: conv_2}, respectively. We control the mesh discretization via the dynamic rate of convergence defined specifically for the finite element simulation of nonlocal elastic models \cite{patnaik2019FEM}. The dynamic rate of convergence is defined as the ratio of the horizon of nonlocality and the size of the discretized mesh element, along a given direction. More specifically, we have the dynamic rate of convergence as $\mathcal{N}_\square=l_f/l_{e_\square}$, where $l_{e_\square},~\square\in\{1,2\}$ denotes the dimension of discretized mesh element along $\hat{e}_\square$. We present the analysis of the numerical convergence for different choices of the dynamic rate of convergence parameter in Fig.~\ref{fig: convergence}. 
We note that for the choice of mesh discretization providing $\mathcal{N}_1 = \mathcal{N}_2 = 10$, the  $\mathbb{L}^1$ norm of the difference between displacements obtained from successive mesh refinements, is less than $1\%$. This implies that a satisfactory convergence of the f-FEM is achieved for the choice of mesh discretization mentioned above. Therefore, this choice of mesh will be retained for the following numerical analyses.

\subsection{Linear elastic response}
\label{sec: linear_results}

\begin{figure*}[b!]
    \centering
    \begin{subfigure}[t]{0.48\textwidth}
        \centering
        \includegraphics[width=\textwidth]{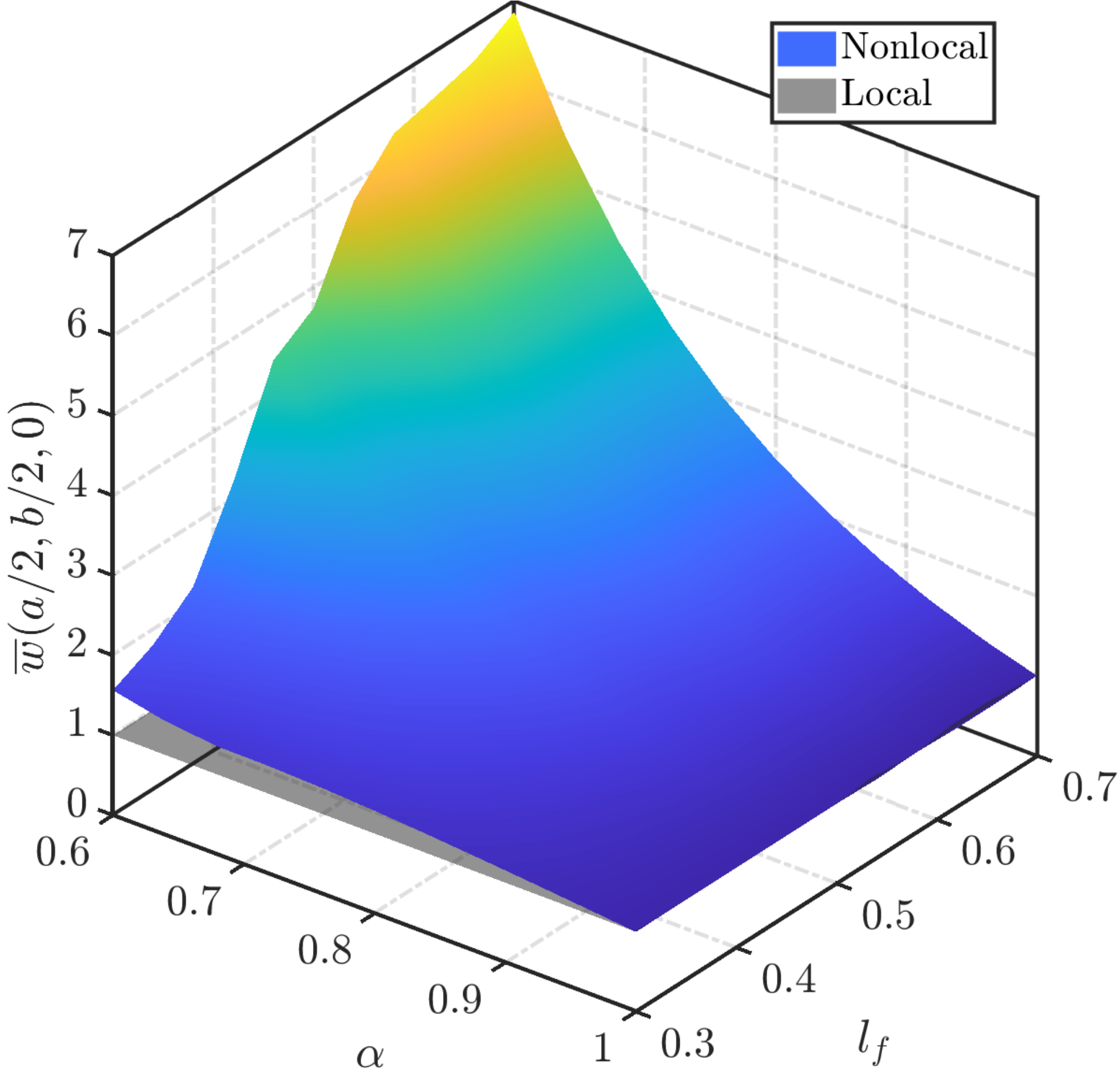}
        \caption{Clamped shell}
        \label{fig: cccc_linear}
    \end{subfigure}
    ~
    \begin{subfigure}[t]{0.48\textwidth}
    \centering
        \includegraphics[width=\textwidth]{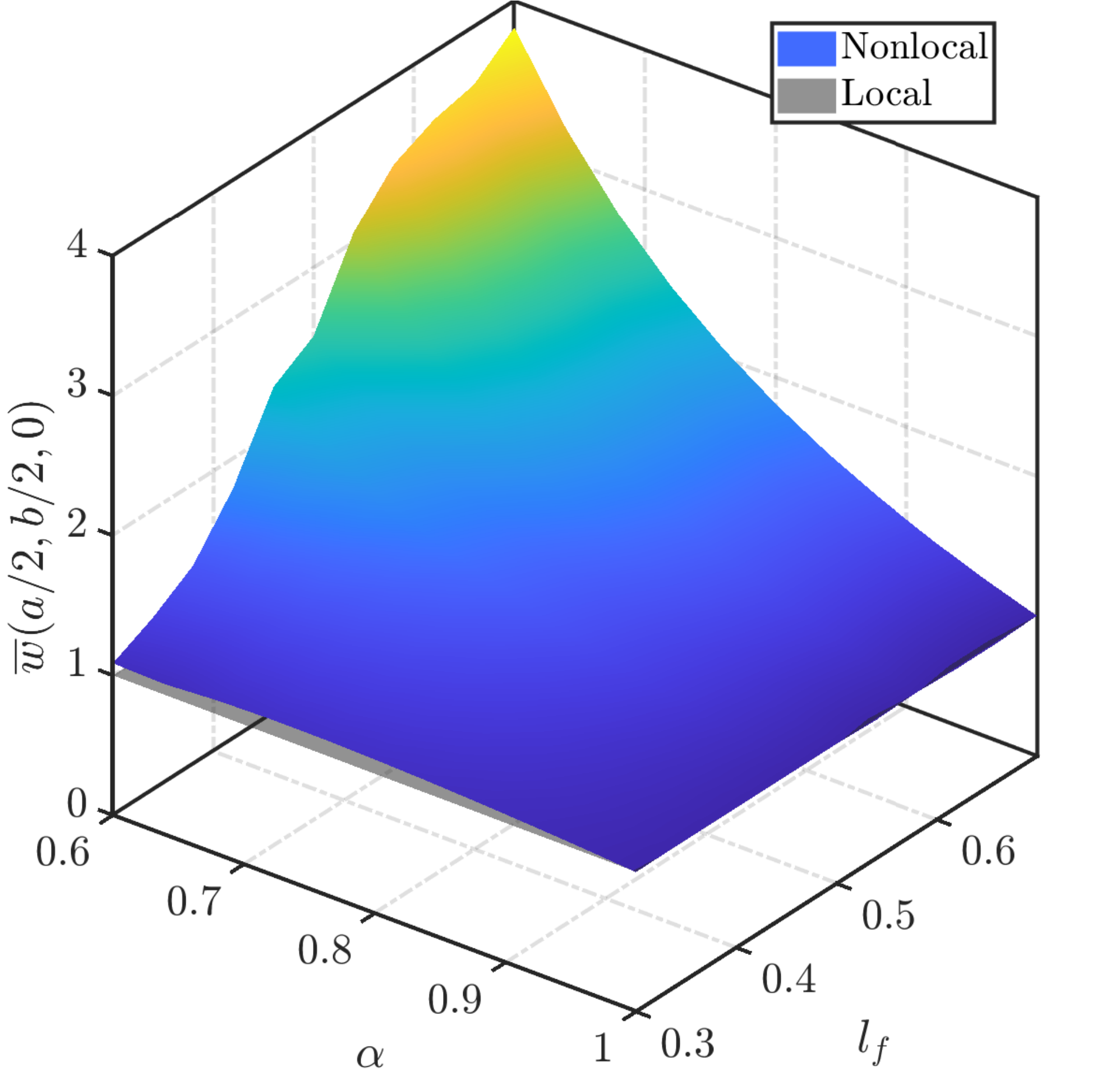}
        \caption{Simply-supported shell}
        \label{fig: ssss_linear}
    \end{subfigure}%
    \caption{Non-dimensionalized transverse displacement at the mid-point and on the mid-plane ($a/2,b/2,0$) of the cylindrical shell with $R/a=10$. Results are compared for different values of the fractional-order constitutive parameters. The non-dimensionalization is achieved following the relation in Eq.~\eqref{eq: disp_nondim} such that local elastic response ($\alpha=1$) is equal to the unity.}
    \label{fig: linear}
\end{figure*}
~
\begin{figure*}[t!]
    \centering
    \begin{subfigure}[t]{0.48\textwidth}
        \centering
        \includegraphics[width=\textwidth]{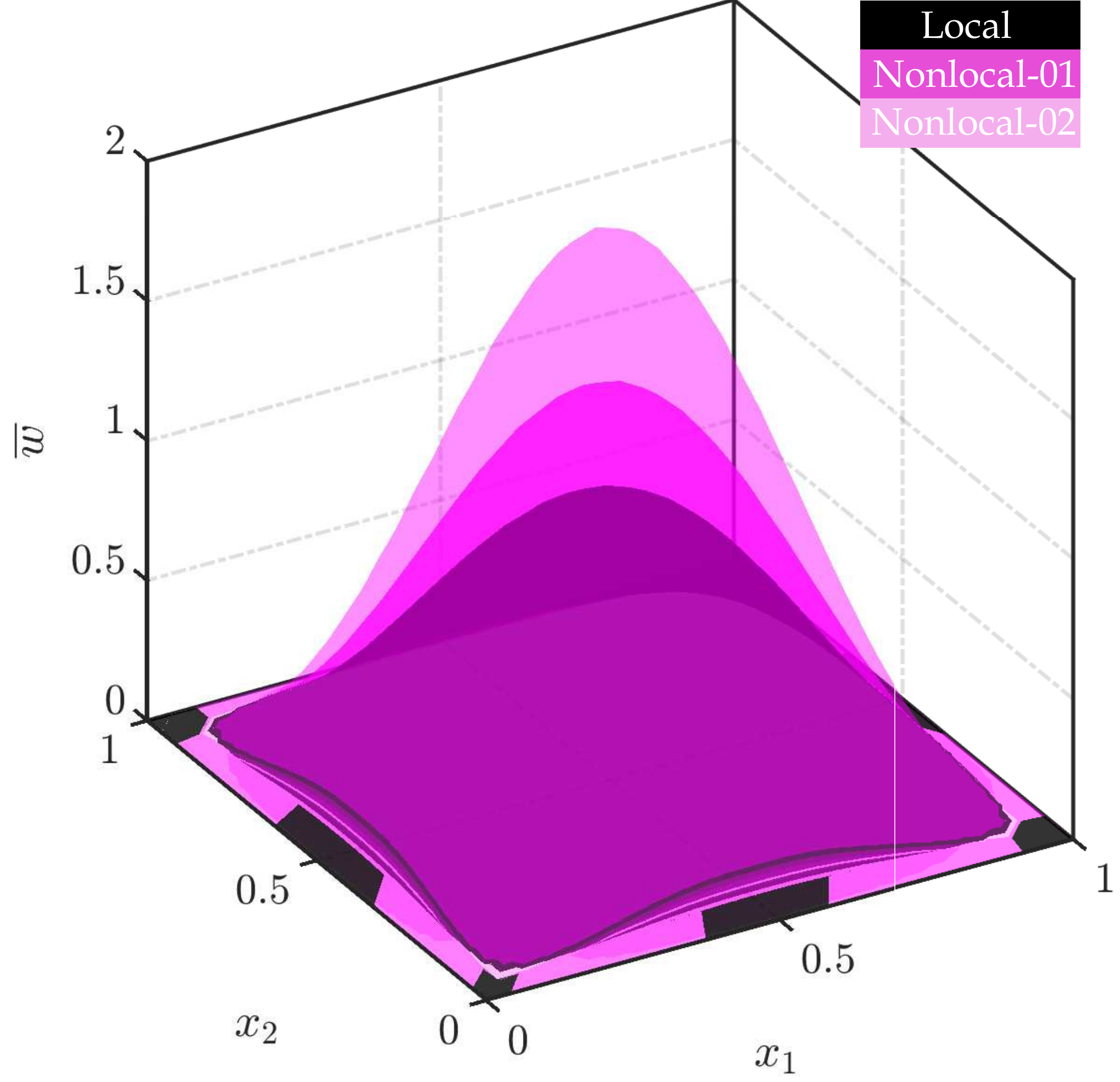}
        \caption{Clamped shell}
        \label{fig: cccc_linear_profile}
    \end{subfigure}
    ~
    \begin{subfigure}[t]{0.48\textwidth}
    \centering
        \includegraphics[width=\textwidth]{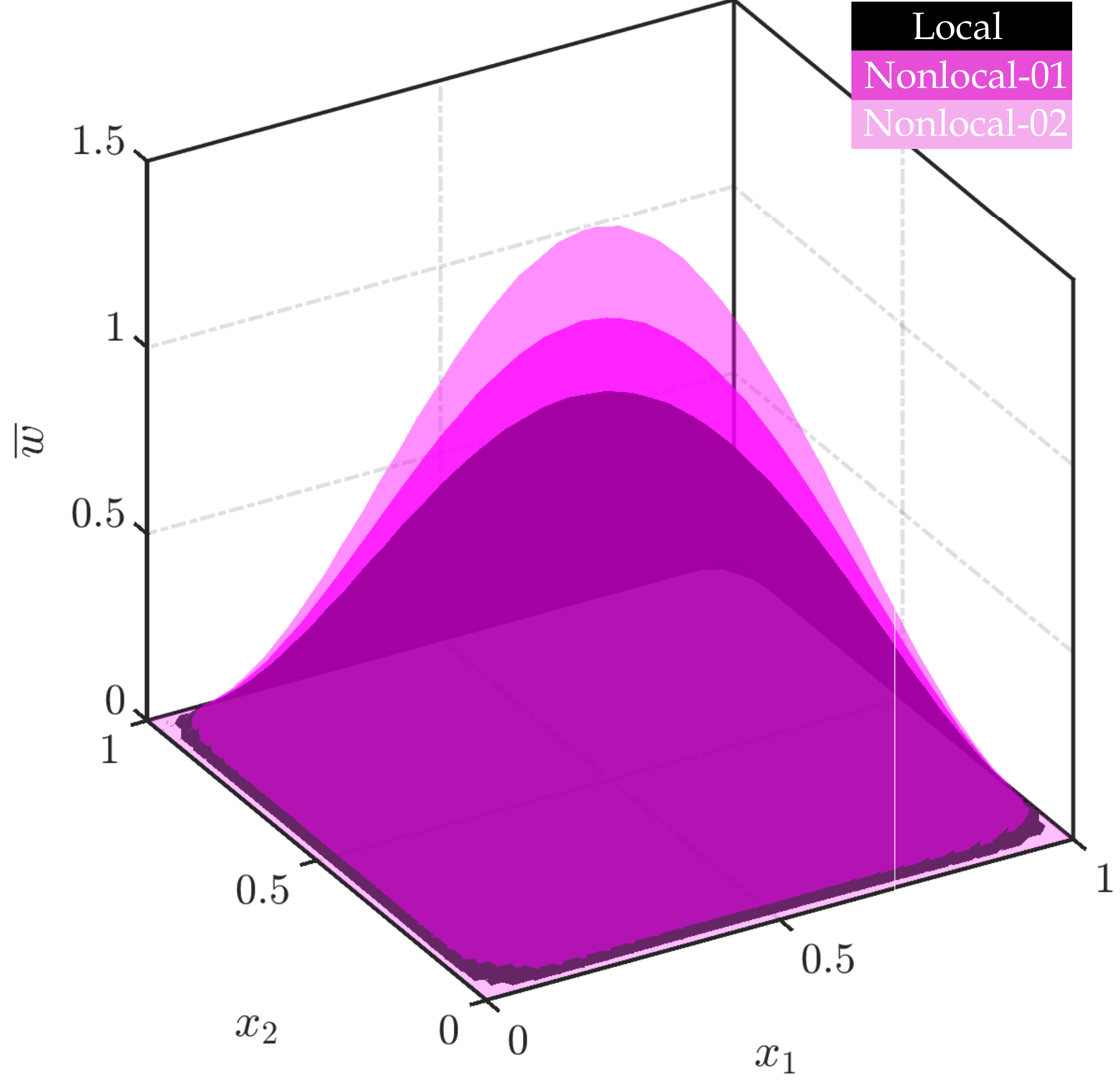}
        \caption{Simply-supported shell}
        \label{fig: ssss_linear_profile}
    \end{subfigure}%
    \caption{Non-dimensionalized displacement field at the mid-plane of the cylindrical panel compared for: (i) local elastic, $\alpha=1.0$; (ii) Nonlocal-01, $\alpha=0.9$, $l_f/a=0.5$; (iii) Nonlocal-02, $\alpha=0.8$, $l_f/a=0.5$}
    \label{fig: linear_profile}
\end{figure*}
In this section, we analyze the linear response of the nonlocal cylindrical panel subject to UDTL and either one of the two different boundary conditions mentioned previously. First, we conduct a parametric study to analyze the effect of the fractional-order constitutive parameters which are, the fractional-order ($\alpha$) and the nonlocal length scale ($l_f$), on the response of the nonlocal panel. The response of the nonlocal panel is presented in Fig.~\ref{fig: linear}, in terms of the maximum transverse displacement which occurs at the mid-point ($a/2,b/2,0$) of the panel. The response of the CCCC and the SSSS panels are provided in Fig.~\ref{fig: cccc_linear} and Fig.~\ref{fig: ssss_linear}, respectively. In each case, the maximum transverse displacement of the the nonlocal panel is non-dimensionalized against the maximum transverse displacement obtained for the local elastic panel, that is:
\begin{equation}
\label{eq: disp_nondim}
    \overline{w}=\frac{w_{\text{nonlocal}}(a/2,b/2,0)}{w_{\text{local}}(a/2,b/2,0)}
\end{equation}
This non-dimensionalization approach clearly highlights the softening that occurs in the panel as a result of the nonlocal interactions. As evident from the results presented in Fig.~\ref{fig: linear}, an increase in the degree of nonlocality (through either a decrease in the value of $\alpha$ or an increase in the value of $l_f$), leads to an increase in the extent of softening, irrespective of the boundary condition. This observation is consistent with the predictions made on the linear elastic behavior of nonlocal beams and plates \cite{patnaik2019FEM}. The consistency in the predictions obtained via the fractional-order shell theory is a direct result of the positive-definite and well-posed nature of the formulation, which is typically not guaranteed through classical integer-order approaches to nonlocal elasticity \cite{patnaik2019FEM,patnaik2022displacement}.

For a more complete analysis, we also compare the deformed shape of the mid-plane of the nonlocal panel with that of the local panel. For this purpose, we consider two different cases: (i) Nonlocal-01: $\alpha=0.9$, $l_f/a=0.5$; and (ii) Nonlocal-02: $\alpha=0.8$, $l_f/a=0.5$. The results of this analysis are presented in Fig.~\ref{fig: linear_profile} and further support the above conclusions, that is the nonlocal interactions soften the panel leading to larger amplitude of displacement compared to the local panel. This behavior is noted irrespective of the choice of the boundary condition (CCCC in Fig.~\ref{fig: cccc_linear_profile} and SSSS in Fig.~\ref{fig: ssss_linear_profile}). Further, for the case Nonlocal-02 where the fractional-order is reduced in order to increase the degree of nonlocality, a higher degree of softening is inferred from the larger amplitude of deformation and independently of the specific boundary conditions.

\subsection{Geometrically nonlinear elastic response}
\begin{figure*}[t!]
    \centering
    \begin{subfigure}[t]{0.48\textwidth}
        \centering
        \includegraphics[width=\textwidth]{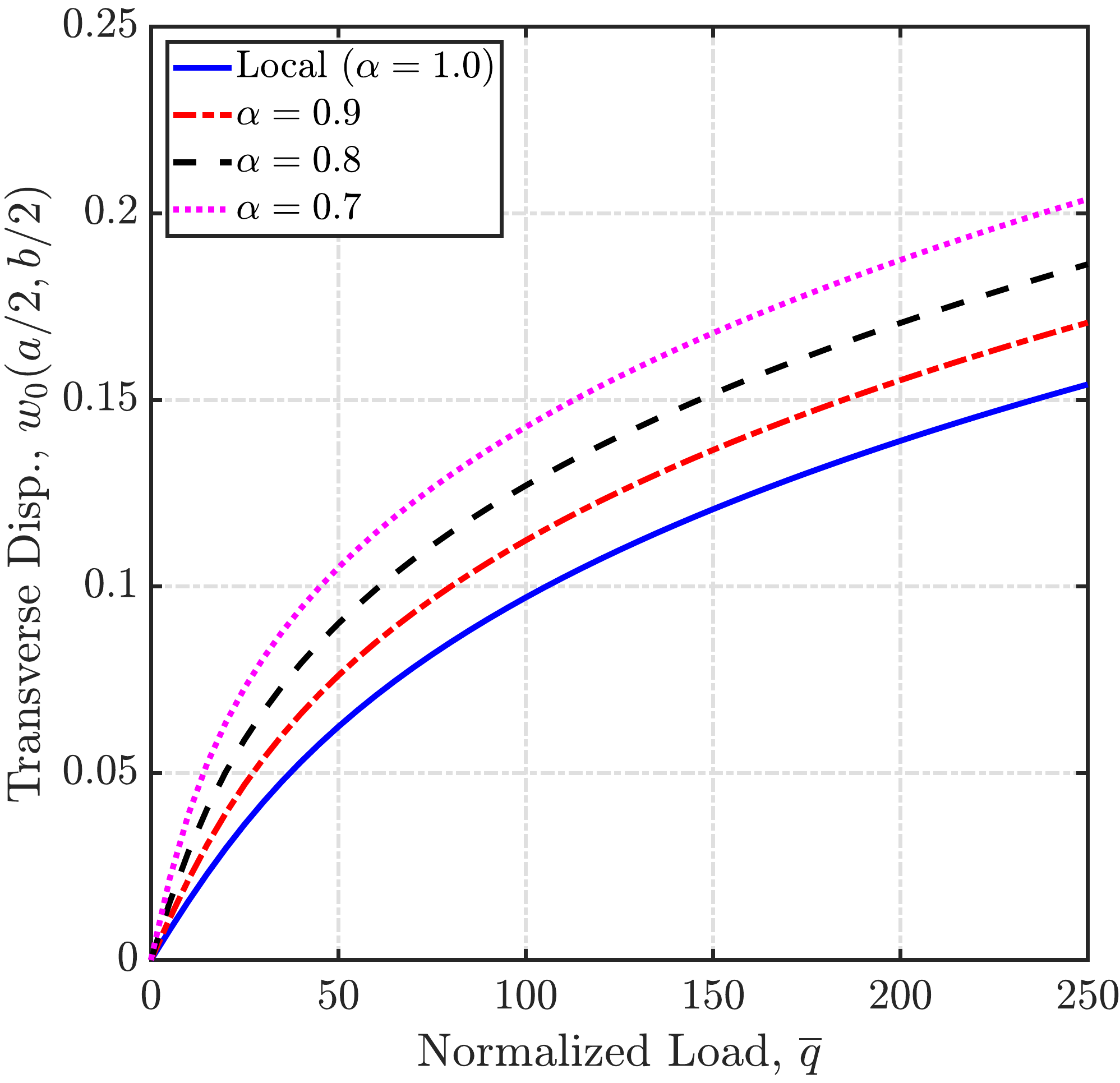}
        \caption{$R/a=10,~~l_f/a=0.5$.}
        \label{fig: CCCC_R10_alpha}
    \end{subfigure}
    ~
    \begin{subfigure}[t]{0.48\textwidth}
        \centering
        \includegraphics[width=\textwidth]{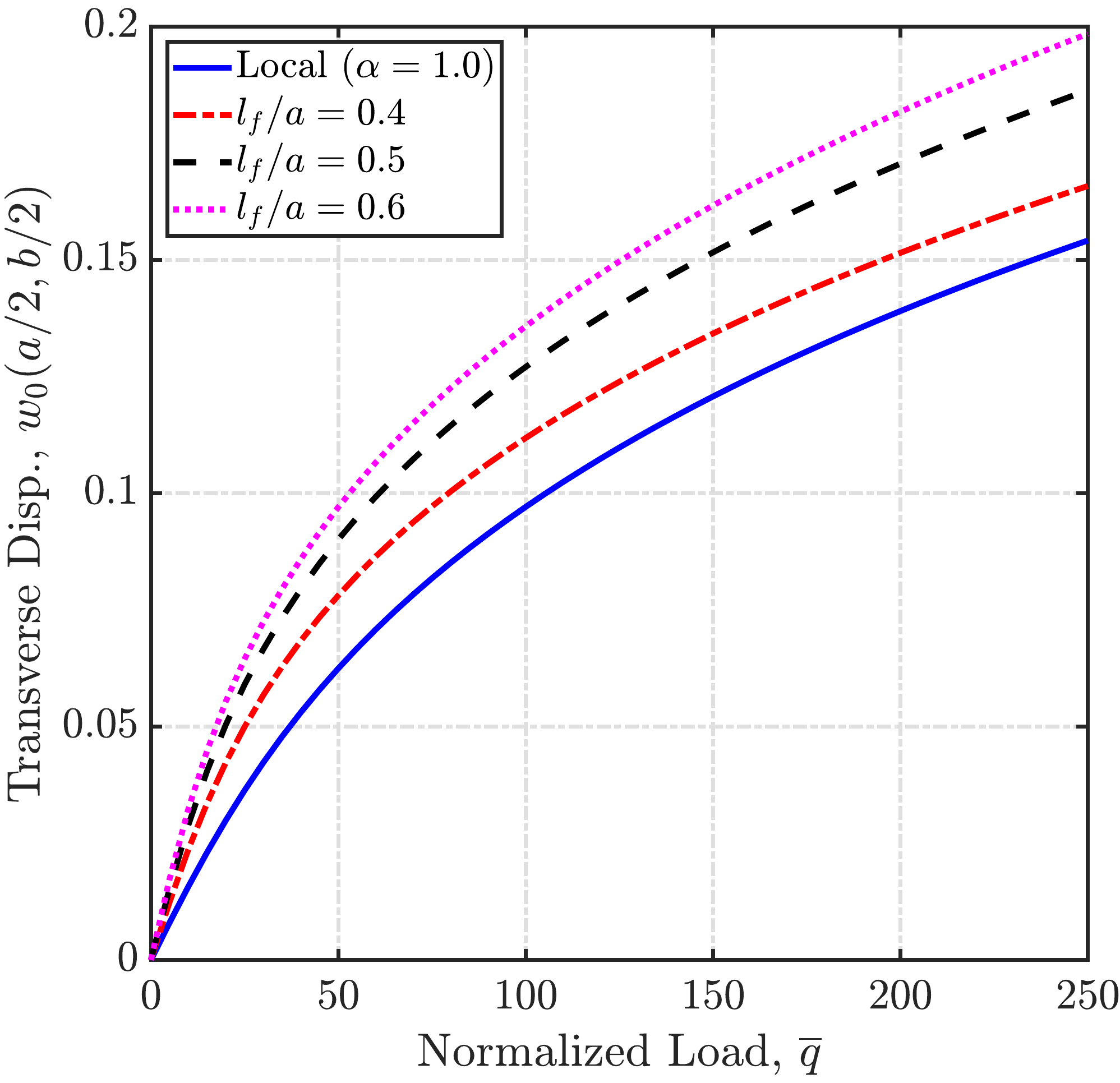}
        \caption{$R/a=10,~~\alpha=0.8$.}
        \label{fig: CCCC_R10_lf}
    \end{subfigure}%
    \caption{Load-displacement curves for the geometrically nonlinear response of clamped cylindrical panels of radius $R/a=10$.  Parametric studies are presented for different values of (a) fractional order, $\alpha$; and (b) length scale $l_f$.} 
    \label{fig: CCCC_nonlinear}
\end{figure*}

\begin{figure*}[b!]
    \centering
    \begin{subfigure}[t]{0.48\textwidth}
        \centering
        \includegraphics[width=\textwidth]{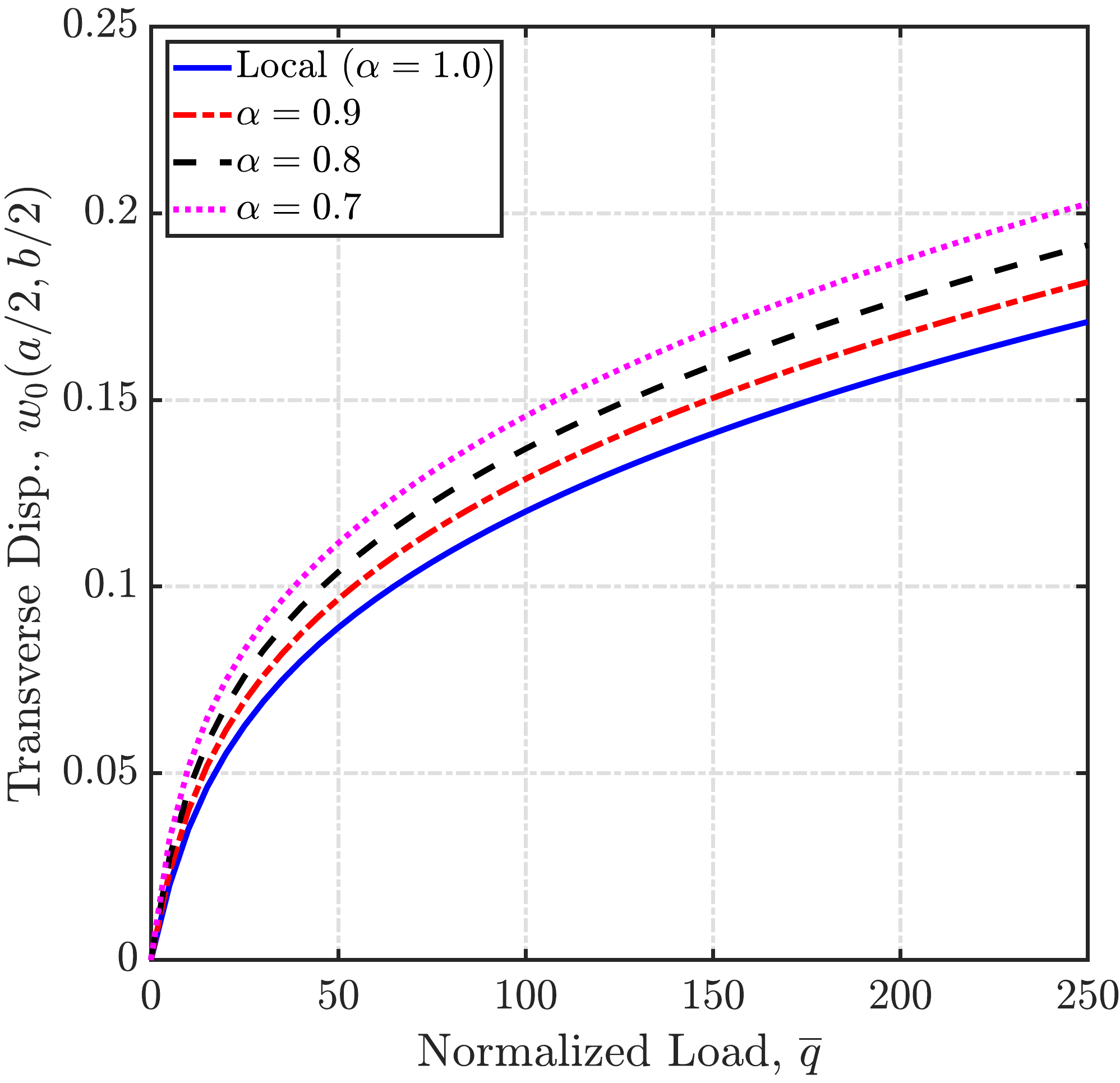}
        \caption{$R/a=10,~~l_f/a=0.5$.}
        \label{fig: SSSS_R10_alpha}
    \end{subfigure}
    ~
    \begin{subfigure}[t]{0.48\textwidth}
        \centering
        \includegraphics[width=\textwidth]{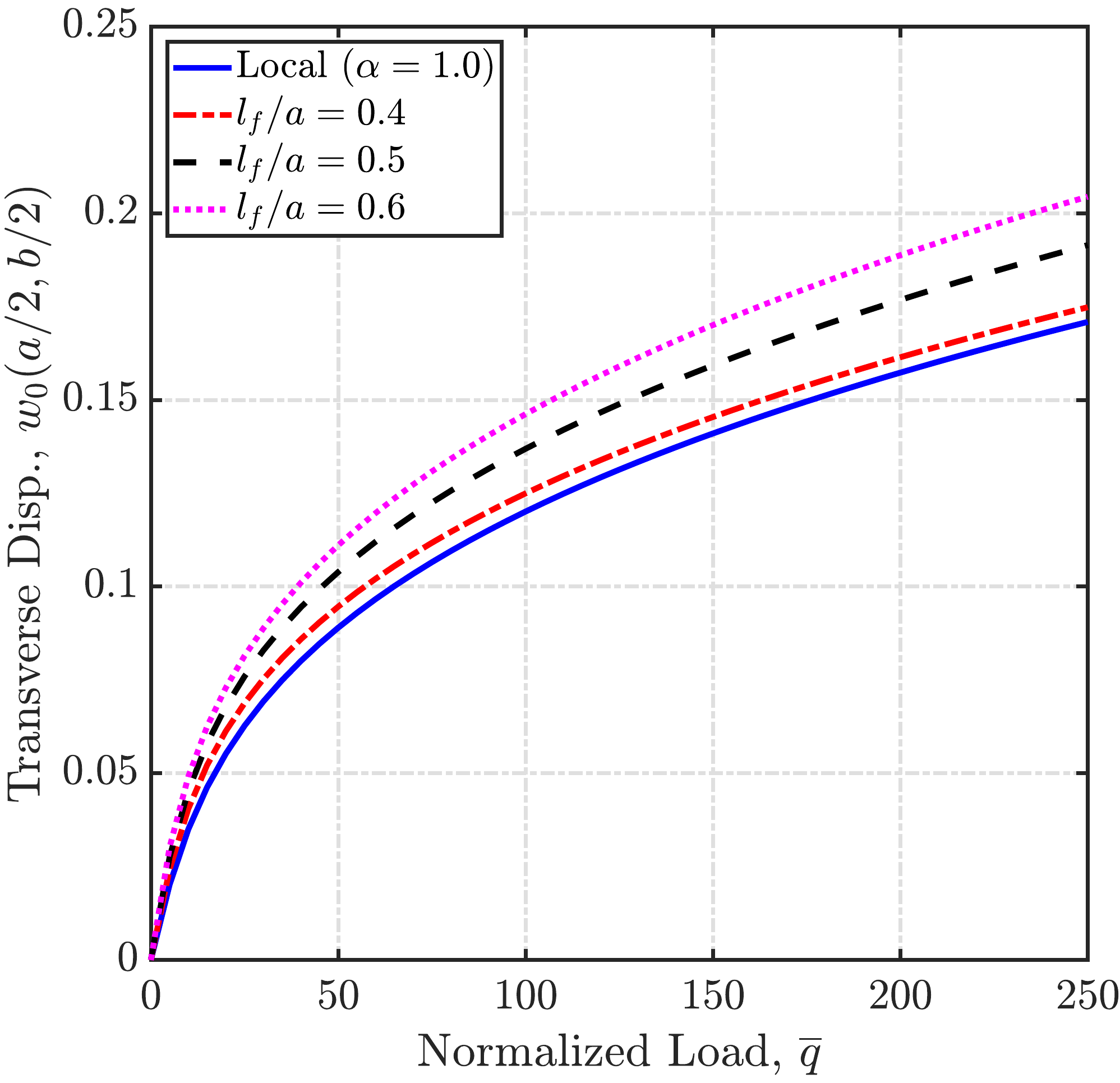}
        \caption{$R/a=10,~~\alpha=0.8$.}
        \label{fig: SSSS_R10_lf}
    \end{subfigure}%
    \caption{Load-displacement curves for the geometrically nonlinear response of simply supported cylindrical panels of radius $R/a=10$. Parametric studies are presented for different values of (a) fractional order, $\alpha$; and (b) length scale $l_f$.} 
    \label{fig: SSSS_nonlinear}
\end{figure*}
In this section, we analyze the geometrically nonlinear response of the nonlocal cylindrical panel using the fractional-order shell theory. The nature of the loading and boundary conditions is identical to that assumed in the linear elastic study. The load-displacement curves for the geometrically nonlinear response of the CCCC panels are provided in Fig.~\ref{fig: CCCC_nonlinear} in terms of the transverse displacement at the center of the mid-plane ($a/2,b/2,0$). The magnitude of the transverse load reported here is non-dimensionalized in the following manner:
\begin{equation}
    \label{eq: load_nondim}
    \overline{q}=q_0\times\left(\frac{L^4}{E\times h^4}\right)
\end{equation}
In Fig.~\ref{fig: CCCC_R10_alpha}, we analyze the influence of the fractional-order $\alpha$ on the nonlinear response, while maintaining the length scale $l_f$ to be constant. Conversely, we analyze the effect of the length scale parameter $l_f$ on the nonlinear response in Fig.~\ref{fig: CCCC_R10_lf}, while maintaining the fractional-order $\alpha$ to be constant. The same analyses are also conducted for SSSS panels, and the corresponding load-displacement curves are provided in Fig.~\ref{fig: SSSS_nonlinear}. 
The effect of geometric nonlinearity is clearly evident from the load-displacement curves for all the cases studied in Figs.~\ref{fig: CCCC_nonlinear} and \ref{fig: SSSS_nonlinear}. In these figures, we observe that an increase in the magnitude of UDTL ($\overline{q}$) results in a nonlinear increment of the maximum value of the transverse displacement for the cylindrical shell. This is unlike a proportional relation between load and displacement expected for linear elastic response. This nonlinearity is demonstrated by local and nonlocal elastic cylindrical shells. In addition, the effect of nonlocality is also manifested in the softening of the nonlinear load-displacement curves in Figs.~\ref{fig: CCCC_nonlinear} and \ref{fig: SSSS_nonlinear}. As evident from the results, the extent of softening is directly related to the degree of nonlocality, that is, an increase in the degree of nonlocality leads to a consistent softening of the panel, irrespective of the nature of the boundary conditions. Further, observe that the softening effect, resulting from the nonlocality, is stronger on the CCCC panels when compared to the SSSS panels. This observation is consistent with predictions made from analogous studies on nonlocal beams and plates \cite{sidhardh2020geometrically,patnaik2020geometrically}, where the softening effect of nonlocality was observed to be stronger on structures subjected to stiffer boundary conditions. The consistency in the nature of predictions, across the different boundary conditions, is in agreement with the linear elastic analysis in \S\ref{sec: linear_results}, and provide evidence of the robustness of the fractional-order continuum theory for the analysis of complex structures exhibiting size-dependent effects. 

\subsection{Influence of the curvature of the panel}
\begin{figure*}[ht]
    \centering
    \begin{subfigure}[t]{0.48\textwidth}
        \centering
        \includegraphics[width=\textwidth]{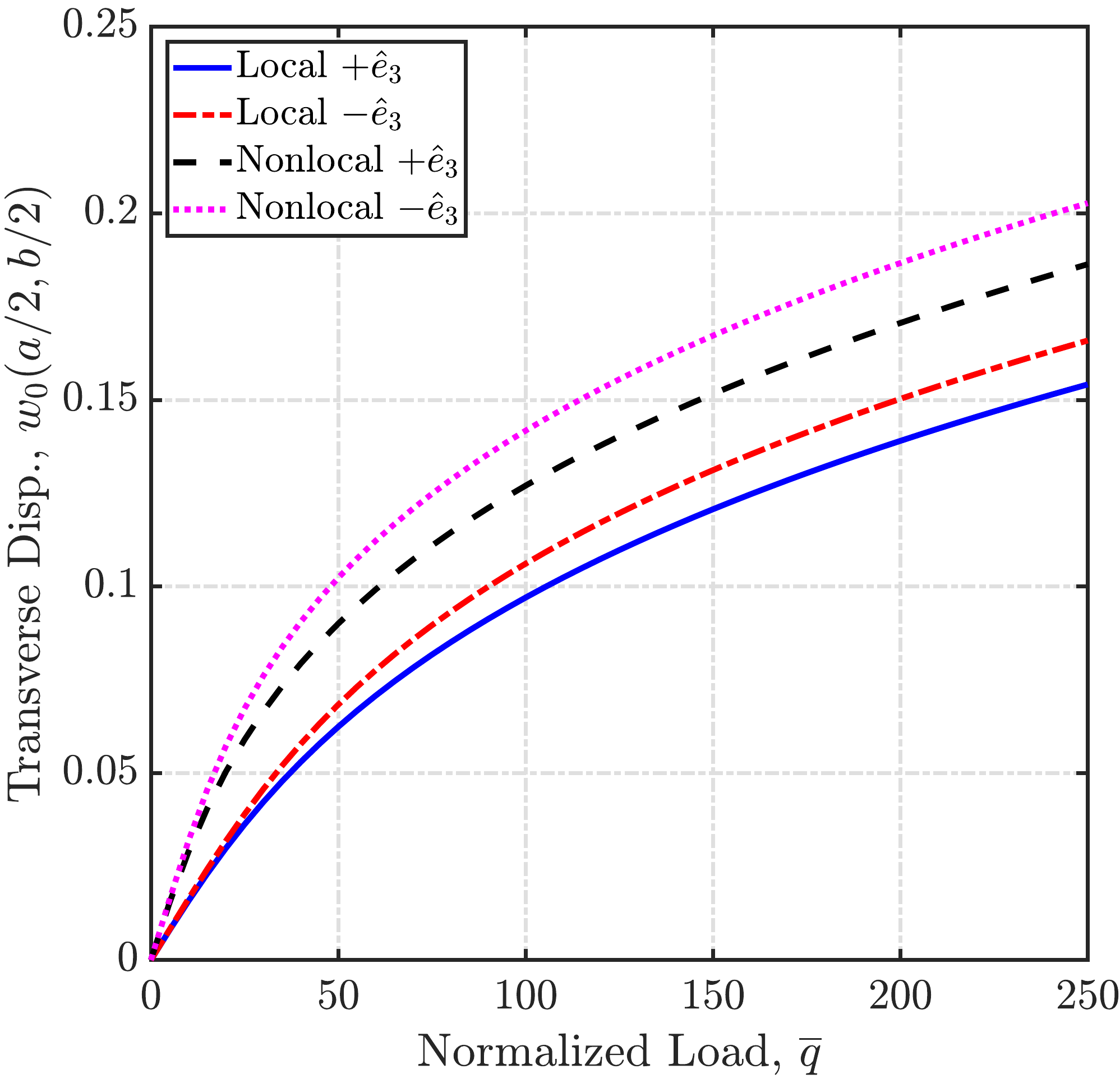}
        \caption{$R/a=10$.}
        \label{fig: R10_pmload}
    \end{subfigure}
    ~
    \begin{subfigure}[t]{0.48\textwidth}
        \centering
        \includegraphics[width=\textwidth]{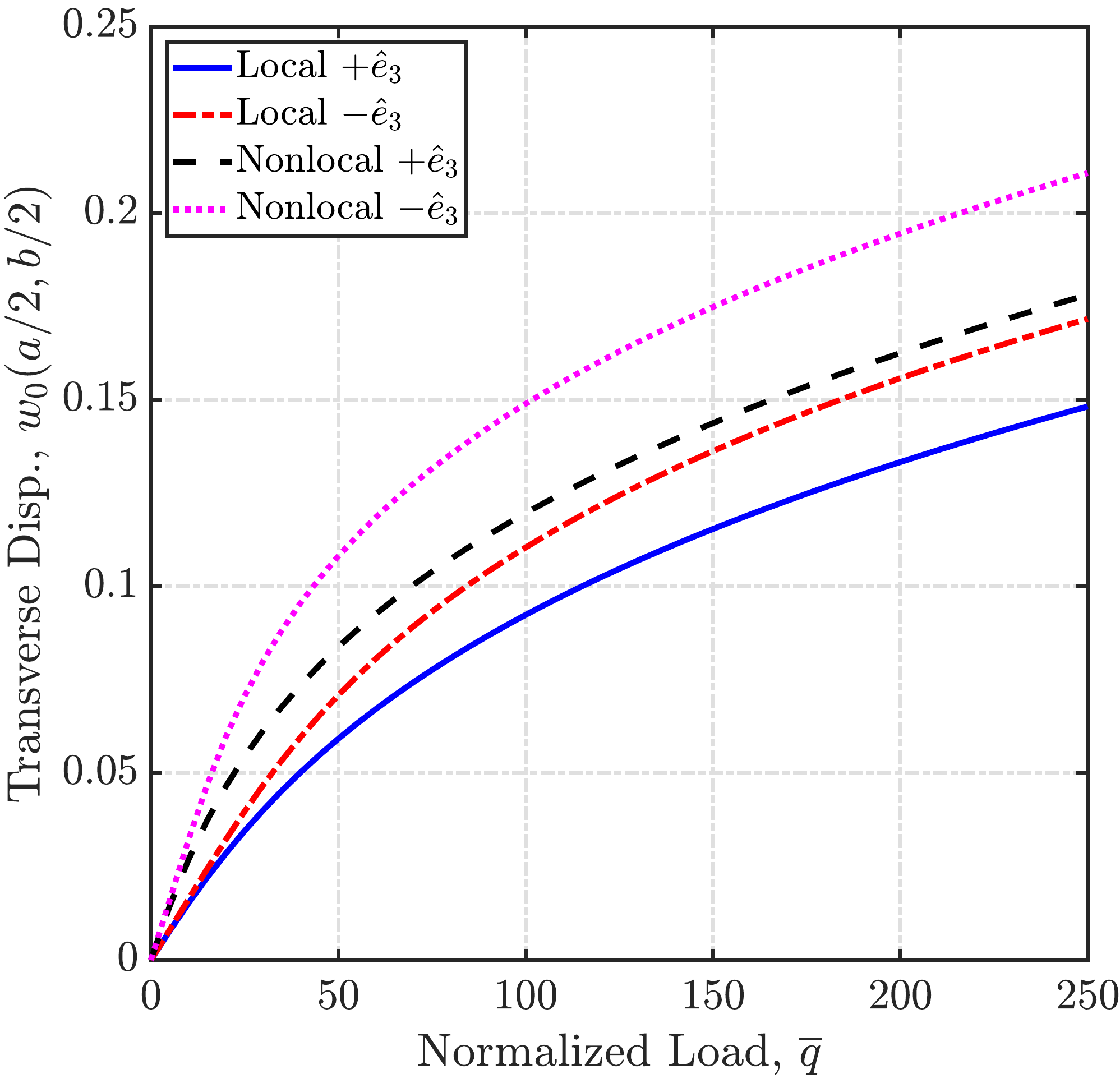}
        \caption{$R/a=5$.}
        \label{fig: R5_pmload}
    \end{subfigure}%
    \caption{Load-displacement curves for the geometrically nonlinear response of clamped cylindrical panels compared for transverse load applied along $+\hat{e}_3$ and $-\hat{e}_3$ directions (see Schematic in Fig.~\ref{fig: schematic}). Local elastic behavior corresponds to $\alpha=1$, and nonlocal elastic behavior is determined for $\alpha=0.8$ and $l_f=0.5$.} 
    \label{fig: pmload}
\end{figure*}

It is well established in the literature that the curvature has a significant and rather complex influence on the stiffness, and consequently, on the nonlinear response, of a cylindrical panel. In fact, when the cylindrical panel is subject to externally applied transverse loads, the direction of the transverse load with respect to the curvature vector, significantly affects the stiffness of the panel, and could possibly lead to the onset of instabilities \cite{reddy2006theory}. In this study, we primarily investigate the interplay between the curvature and the nonlocality and their effect on the overall stiffness and on the nonlinear response of the panel. For this purpose, we compare the nonlinear load-displacement curves of a local and nonlocal CCCC panel subjected to UDTL along $\pm \hat{e}_3$ directions. Note that, for the $+\hat{e}_3$ direction, the UDTL is along (or, equivalently, parallel to) the radius of curvature of the panel (see Fig.~(\ref{fig: schematic}). Similarly, for the $-\hat{e}_3$ direction, the UDTL is opposite (or, equivalently, anti-parallel) to the radius of curvature of the panel. The results of this study are provided in Fig.~\ref{fig: pmload} for two different radii of curvature of the cylindrical panel. Note that the strength of nonlocality and the radius of curvature, for all the cases considered in the study, are chosen such that the nonlinear response of the panel is stable for all values of the UDTL.

As evident from the results, the panel subject to the UDTL of a given magnitude along the $+\hat{e}_3$ exhibits a stiffer response when compared to the panel subjected to a UDTL of the same magnitude along the $-\hat{e}_3$-direction. From a physical perspective, the transverse load applied along $+\hat{e}_3$ (see Fig.~\ref{fig: schematic}) induces tensile in-plane stresses within the panel and hence, increases the nonlinear stiffness of the panel when compared to a plate ($R\to \infty$) with same in-plane dimensions. In contrast, the transverse load along $-\hat{e}_3$ induces compressive in-plane stresses and results in a reduction of the nonlinear stiffness of the panel, when compared to a plate. Note that the stiffer response of the panel to the transverse load applied along the radius of curvature (that is, $+\hat{e}_3$) is observed irrespective of the degree of nonlocality. However, the difference in the degree of softening of the nonlocal shell is higher in the case of UDTL applied in a direction opposite to the radius of curvature (that is, $-\hat{e}_3$), and increases with the increasing curvature (or, equivalently, with the decreasing radius $R$). Hence, it appears that the curvature and the degree of nonlocality of the panel, together play a complex role in determining a stiffening or softening of the panel response. 

\section{Conclusions}
This study extended the fractional-order formulation, previously developed for beams and plates, to the analysis of nonlocal cylindrical shell panels. By  developing geometrically nonlinear fractional-order kinematic relations for cylindrical shells, we obtained a frame-invariant fractional-order framework that is mathematically, physically, and thermodynamically consistent; in contrast to the already existing integer-order models.
%
%
The fractional-order theory is used as the foundation to develop 2D numerical models of both the linear and the geometrically nonlinear response of nonlocal shells, based on first-order shear displacement theory. The models are solved numerically via the fractional finite element method for a variety of test cases. Results highlight the emergence of a softening behavior caused by the nonlocal interactions, irrespective of the type of boundary conditions. This consistency of the response across different loading and boundary conditions overcomes an important issue observed in existing methodologies for the analysis of nonlocal solids based on integer-order models. Our proposed approach also allowed establishing an important feature characteristic of the nonlocal response of shell structures, that is the degree of softening due to the nonlocal interactions increases with the increasing curvature. The understanding and modeling of the role that the curvature plays on the degree of softening is expected to be of significance for studies involving instabilities of cylindrical panels. Finally, we note that the present analysis framework for nonlocal shells is expected to form the basis for further studies on nonlocal curved structures such as those involving layered and porous media, or even biological materials like tissues and bones. The models developed here based on the fractional-order approach are expected to be used in the analysis of long-range interactions (nonlocal), spatial multiscale effects, and complex heterogeneities, on the structural response.\\

\noindent \textbf{Acknowledgements:} The authors gratefully acknowledge the financial support of the Defense Advanced Research Project Agency (DARPA) under grant \#D19AP00052, and of the National Science Foundation (NSF) under grants MOMS \#1761423 and DCSD \#1825837. S.P. acknowledges the financial support of the Hugh W. and Edna M. Donnan Fellowship awarded by the School of Mechanical Engineering at Purdue University. Any opinions, findings, and conclusions or recommendations expressed in this material are those of the author(s) and do not necessarily reflect the views of the National Science Foundation. The content and information presented in this manuscript do not necessarily reflect the position or the policy of the government. The material is approved for public release; distribution is unlimited.\\

\noindent \textbf{Competing interests:} The authors declare no competing interest.

\bibliographystyle{naturemag}
\bibliography{References}
\end{document}